\documentclass[USenglish,oneside,twocolumn]{article} %
\usepackage[utf8]{inputenc}%
\usepackage[big]{dgruyter_NEW}
\usepackage{xspace}
\usepackage{comment}
\usepackage{graphicx}
\usepackage{amssymb, amsthm, amsfonts, latexsym}
\usepackage{multirow}
\usepackage{subcaption}
\usepackage{wasysym}
\usepackage{pifont}
\usepackage{balance}
\usepackage{siunitx}
\usepackage{color}
\usepackage{mathtools}
\usepackage{hyperref}
\usepackage{tabularx}
\usepackage[scr=boondox]{mathalfa}

\newcommand{\eg}{\emph{e.g.}\xspace}

\newcommand{\ie}{\emph{i.e.}\xspace}
\newcommand{\etc}{\emph{etc.}\xspace}
\newcommand{\etal}{{\em et~al.}\xspace}
\newcommand{\sysname}{Tempest\xspace}

\newcommand\commanum[1]{\num[group-separator={,}, group-minimum-digits=4]{#1}}

\newcommand{\IQR}{\text{IQR}}

\DeclarePairedDelimiter\abs{\lvert}{\rvert}

\newcommand{\squishlist}{
 \begin{list}{$\bullet$}
  { \setlength{\itemsep}{0pt}
     \setlength{\parsep}{3pt}
     \setlength{\topsep}{3pt}
     \setlength{\partopsep}{0pt}
     \setlength{\leftmargin}{.1em}
     \setlength{\labelwidth}{1em}
     \setlength{\labelsep}{0.5em} } }

\newcommand{\squishend}{
  \end{list}  }

\title{\sysname: Temporal Dynamics in Anonymity Systems}
\author*[1]{Ryan Wails}
\author[2]{Yixin Sun}
\author[3]{Aaron Johnson}
\author[2]{Mung Chiang}
\author[2]{Prateek Mittal}
\affil[1]{U.S. Naval Research Laboratory, E-mail: ryan.wails@nrl.navy.mail}
\affil[2]{Princeton University, E-mail: \{yixins, chiangm, pmittal\}@princeton.edu}
\affil[3]{U.S. Naval Research Laboratory, E-mail: aaron.m.johnson@nrl.navy.mail}
\cclogo{\includegraphics{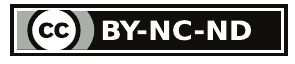}}
\runningtitle{\sysname: Temporal Dynamics in Anonymity Systems}
\journalname{Proceedings on Privacy Enhancing Technologies}

\begin{document}

\begin{abstract}{
Many recent proposals for anonymous communication omit from 
their security analyses a consideration of 
the effects of \emph{time} on important system components.  
In practice, many components of anonymity systems, such as the client location and
network structure, exhibit changes and patterns over time.
In this paper, we focus on the effect of such \emph{temporal dynamics} on the security
of anonymity networks. We present \sysname, a suite of novel attacks based on 
(1) client mobility, (2) usage patterns, and (3) changes in the underlying network routing. 
Using experimental analysis on real-world datasets, we demonstrate that these
temporal attacks degrade user privacy across a wide range of anonymity networks, 
including deployed systems such as Tor; path-selection protocols for Tor such as DeNASA, TAPS, and 
Counter-RAPTOR; and network-layer anonymity protocols for Internet routing such as
Dovetail and HORNET. The degradation is in some cases surprisingly severe. For example, a single 
host failure or network route change could quickly and with high certainty identify the client's ISP
to a malicious host or ISP.
The adversary behind each attack is relatively weak --- generally passive and
in control of one network location or a small number of hosts.
Our findings suggest that designers of anonymity systems should rigorously consider the impact of
temporal dynamics when analyzing anonymity.}
\end{abstract}

\maketitle

\vspace*{-.3in}

\vspace*{-.4in}
\section{Introduction}

\vspace*{-.15in}
Anonymous communication is a key privacy-enhancing technology that aims to 
protect user identity in online communications~\cite{tor-design,onion-routing-ih96,freedom2-arch,web-mix-pet2000,pets2011-i2p}. 
The most widely-used 
anonymity protocol today is onion routing~\cite{onion-routing-ih96}, 
which in the form of the Tor network~\cite{tor-design} is 
estimated to have over 2 million users a day. The Tor network 
comprises over \commanum{7000} volunteer proxies, carries 100 Gbps 
of traffic, and is widely used by citizens, journalists, whistleblowers, 
businesses, governments, law-enforcement, and intelligence agencies
\cite{torproject,tormetrics}.
An important thread of research has proposed 
new anonymity systems that improve on Tor in the context of network-level 
adversaries, such as Autonomous Systems (ASes) that have vast visibility 
into Internet traffic. Systems such as DeNASA, Astoria, TAPS, and 
Counter-RAPTOR have modified path-selection algorithms for onion routing 
to mitigate the threat of AS-level adversaries~\cite{denasa-pets2016,astoria-ndss2016,taps-ndss2017,counterraptor-sp2017}. 
Systems such as LAP, Dovetail, PHI, and HORNET have moved cryptographic 
functionality for anonymous communication from end hosts into the Internet 
routing infrastructure to improve performance~\cite{oakland2012-lap,pets14-dovetail,hornet-ccs2015,phi-popets17}.

However, for simplicity of security analysis, designers of these systems
abstract away important components of the system, which could impact user 
anonymity in practice. In particular, one simplification commonly used in the analysis 
of anonymity protocols is to \emph{limit the effects of time on the operation 
of the protocol}~\cite{lastor,astoria-ndss2016,denasa-pets2016,oakland2012-lap,pets14-dovetail}. 
For example, security analyses typically assume that each user 
communicates with a fixed destination once, that the set of participants 
in the protocol is static, or that the network structure is static. 
The question then arises: \emph{what are the effects 
of the temporal dimension of system operations on user anonymity?}

{\bf Contributions.} In this paper, we present Tempest: a set of attacks that demonstrates the
impact of temporal dynamics on the security of several prominent anonymity protocols.  
We target Tor and some of the latest proposals for improving its security against AS-level adversaries
(namely, DeNASA~\cite{denasa-pets2016}, TAPS~\cite{taps-ndss2017}, and
Counter-RAPTOR~\cite{counterraptor-sp2017}),
as Tor has 
proven to be the most popular protocol for the current Internet. 
We also target proposals for network-layer anonymity that 
represent the main ideas for providing anonymity against AS-level adversaries 
in a next-generation Internet (namely, Dovetail~\cite{pets14-dovetail} and
HORNET~\cite{hornet-ccs2015}). 

\begin{table*}[!ht]
\centering
\caption{This paper identifies and analyzes temporal dynamics (\sysname attacks) that degrade user privacy in anonymity systems, including 
Tor, proposals for improving path selection in Tor (top half), and network-layer anonymity protocols (bottom half). 
}
\begin{tabularx}{\textwidth}{l|XXX}
 & Exploiting Client Mobility & Exploiting User Behavior & Exploiting Routing Changes\\
\hline
Vanilla Tor~\cite{tor-design}    & Novel (\S\ref{subsec:tor}) & Known \cite{borisov:ccs07, latencyleak-tissec, ccs2013-usersrouted, onions-target-wpes17}  & Known \cite{raptor-usenix2015}   \\
DeNASA~\cite{denasa-pets2016}    & Novel (\S\ref{app:mobility-denasa})    & Novel (\S\ref{subsec:denasa_multiple_conn})                     &           \\
Counter-RAPTOR~\cite{counterraptor-sp2017} & Novel (\S\ref{subsec:counter-raptor})  & Novel (Resistance, \S\ref{app:user-counter-raptor}) & Known (Resistance) \cite{counterraptor-sp2017} \\ 
TAPS~\cite{taps-ndss2017}        &                                  & Known (Resistance) \cite{taps-ndss2017}      & Novel (\S\ref{subsec:taps})     \\
Astoria~\cite{astoria-ndss2016}  &                                  &  Known \cite{taps-ndss2017}        &             \\

\hline
HORNET~\cite{hornet-ccs2015}         & Novel (\S\ref{subsec:hornet_mobility})                &                     & Novel (\S\ref{subsec:hornet_routing})               \\
Dovetail~\cite{pets14-dovetail}         &  	       & Novel (\S\ref{subsec:dovetail_multiple_conn})          &                  \\
PHI~\cite{phi-popets17}            &                      & Novel (\S\ref{app:user-phi})                     &                     \\ 
LAP~\cite{oakland2012-lap}         & Novel (implied by \S\ref{subsec:hornet_mobility})      &                        & Novel (implied by \S\ref{subsec:hornet_routing})      \\
\end{tabularx}
\label{tab:overview}
\vspace{-.2in}
\end{table*}

We consider the vulnerability of such protocols to deanonymization due to 
the effects on anonymous-communication paths of three main types of temporal dynamics: 
(1) client mobility, (2) user behavior over multiple connections, and
(3) network routing dynamics. We consider especially a \emph{patient} adversary
that is interested in performing \emph{long-term} attacks on 
anonymous communication. Such an adversary is a real concern of today's Tor
users~\cite{ccs2013-usersrouted,oft-target-hotpets2017}, for example those
avoiding mass surveillance.
We propose and, using real-world datasets, evaluate attacks that allow an
adversary to exploit long-term observations about \emph{anonymity path changes}
for deanonymizing user identities.

We find that Tempest attacks have significant impact on the anonymity 
of these systems. One impact that we show is that, in Tor-based systems, path changes due to client
mobility allow an increasing number of AS-level adversaries to observe client traffic, 
compromising client identity with a degree much greater than previously thought possible. 
In network-level anonymity systems, adversaries can also correlate partial 
information about anonymity paths with auxiliary information about client 
movements to deanonymize client identity. Our work 
presents the first analysis of the impact of path changes due to client mobility. 
Another impact that we show is that path changes due to user behavior,
such as \emph{multiple} connections to the same destination at different times,
and path changes due to network routing updates allow an adversary 
to \emph{combine} probabilistic information leaks inherent in path-selection algorithms 
to deanonymize clients' ASes over time. Note that inferring the AS of a client represents 
a significant reduction in client anonymity (the typical anonymity set comprises over 
50,000 ASes without our attacks). Our work is the 
first demonstration of how probabilistic information leaks due to the restricted AS-level 
Internet topology can be aggregated over time.

Our results present a new evaluation paradigm for important classes of 
anonymous-communication protocols. They suggest that designers of anonymity 
systems should thoroughly consider the impact of temporal dynamics when
analyzing system security. Our work further motivates the design 
of anonymous communication protocols that are resilient when used over time
and under changing circumstances.
\vspace*{-.3in}
\section{Overview of Tempest Attacks}
\vspace*{-.15in}
In this section, we provide an overview of Tempest attacks and summarize 
our key findings. 

{\bf Exploiting Client Mobility.} We demonstrate how an adversary can exploit 
information leakage via naturally-occurring real-world movements of clients. Client 
mobility results in connections to anonymity networks appearing from 
different network locations over time; we find that this enhances an adversary's 
ability to perform traffic-analysis attacks and deanonymize client communications. 
We experimentally quantify the degradation in anonymity for Vanilla Tor
(\ie{} plain Tor as it exists today), 
Counter-RAPTOR~\cite{counterraptor-sp2017}, and HORNET~\cite{hornet-ccs2015} using 
 real-world location datasets to model client mobility. Across all studied systems, 
we find that considering the effects of client mobility results in an
order-of-magnitude degradation in client anonymity: (1) for Vanilla Tor, client mobility 
increases the exposure of the client-Tor communications to
AS-level adversaries, due to heterogeneous network paths originating from 
varying client locations; (2) for Counter-RAPTOR, client mobility increases 
an adversary's ability to actively manipulate BGP routing and hijack/intercept 
client traffic to the anonymity network, due to a fundamental mismatch in assumptions 
between its location-aware path selection and the dynamics of client mobility; and 
(3) for HORNET, client mobility results in changes in the network paths between 
a client and its destination over time, which can be correlated with external
(non-anonymous) location datasets to deanonymize the anonymity-network connections.

{\bf Exploiting User Behavior.} We consider users that regularly connect to
a destination and demonstrate how an adversary can take advantage of that
to deanonymize users in several prominent anonymity proposals. 
Multiple user connections allow an adversary to aggregate probabilistic information 
leakage from connections over time and eventually deanonymize the user
by identifying his AS.
We experimentally quantify this degradation in anonymity over multiple connections for
DeNASA~\cite{denasa-pets2016}, 
a location-aware path selection algorithm for Tor that avoids suspect ASes,
and Dovetail~\cite{pets14-dovetail}, a network-level anonymity protocol that uses 
a level of indirection within Internet communications to provide anonymity. We find 
that multiple connections have devastating consequences on user anonymity in DeNASA 
and Dovetail. The path selection algorithms in both DeNASA (focusing on the 
first-hop/guard relay) and Dovetail leak partial information about the client's 
network location, leading to a continual reduction in anonymity as the client makes
connections. The speed of this reduction is surprisingly fast for some
unfortunate clients.

{\bf Exploiting Routing Changes.} We show how an adversary can exploit
naturally-occurring routing changes to compromise client anonymity. Similar to the impact of
client mobility, routing changes leak additional information to an adversary as they occur,
which can be aggregated over time to make accurate inferences about client location.
We experimentally quantify the degradation in anonymity due to routing changes for TAPS~\cite{taps-ndss2017}, 
a trust-aware path selection algorithm for Tor, and HORNET~\cite{hornet-ccs2015}. For both 
TAPS and HORNET, routing changes lead to varying anonymity sets for clients over time, 
allowing an adversary to intersect the anonymity sets at different points in time and infer client 
network locations (client ASes).

{\bf Impact.} The impact of the \sysname attacks is summarized in Table~\ref{tab:overview}. For each
of the three temporal dynamics considered, we demonstrate attacks that weaken security in at least
one onion-routing protocol and one network-layer anonymity protocol. We note that our results on
exploiting client mobility represent the first analysis of this issue, to the best of our knowledge,
and we demonstrate its negative impact on a range of systems. We also note that our results on
exploiting user behavior and routing changes include several novel attacks across recently-proposed
onion-routing and network-layer anonymity protocols, suggesting a significant re-evaluation of their
effectiveness. In particular, our work is the first to consider 
the aggregation of probabilistic information leaks due to the restricted AS-level Internet topology 
over time.

To highlight the broad impact of \sysname, Table~\ref{tab:overview} includes novel attacks 
that appear in the Appendix, including exploiting client mobility in DeNASA, exploiting
user behavior in Counter-RAPTOR, and exploiting user behavior in PHI (note that the results in
the main body of the paper are self-contained).
Table~\ref{tab:overview} also includes entries for protocols not studied directly in this work. It places
into context an attack~\cite{taps-ndss2017} on Astoria~\cite{astoria-ndss2016} that is similar to
our attacks exploiting user-behavior dynamics. It also indicates that some of our attacks should
also be effective against LAP~\cite{oakland2012-lap}, as LAP
reveals strictly more to the adversary about the source
and destination of a connection than HORNET does.

Due to the variety of systems considered, the \sysname attacks 
vary in the kind of anonymity degradation they achieve and in the 
adversary capabilities that they require. In several cases we attack 
anonymity using the same notions and metrics used to argue for the
system's effectiveness by its designers. In every case, the adversaries
we consider fall within the threat model stated for 
the system under analysis. In addition, the adversaries we consider are
generally passive and need to control only one or a small number of network entities
(\eg{} an AS, a website, or a Tor relay). To summarize our contributions, we identify
the effects of temporal dynamics on paths
in anonymity systems as a general concern affecting the anonymity of 
those systems. We present the \sysname attacks and show that they 
can reduce the anonymity of Tor, suggested Tor improvements, and 
network-layer anonymity protocols.

\vspace*{-.25in}
\section{Background} \label{sec:background}

\vspace*{-.15in}
In this section, we present the required background on Internet routing and anonymity protocols.

{\bf Network Routing.} \label{subsubsec:network-routing}
Routing in the Internet is set up among routers via the Border Gateway Protocol
(BGP). BGP produces routing paths between the autonomous subnetworks that
comprise the Internet, called \emph{Autonomous Systems} (ASes). There are
roughly \commanum{58000} ASes advertising routes on the
Internet~\cite{cidr-report}. Each AS is connected to at least one other AS, and the connected ASes exchange traffic with each other in a variety of bilateral relationships that specify when traffic should be sent and how it is paid for. 
In BGP, routing operates on variable-length \emph{IP prefixes}, which are each a sequence of bits that is compared to beginning of the destination IP address to route a packet.

{\bf Onion Routing.} \label{subsubsec:onion-routing}
Onion routing~\cite{onion-routing-ih96} achieves anonymous communication online by encrypting the
network traffic and sending it through a sequence of \emph{relays} before going to the
destination. The relays run at the application layer on the hosts, and traffic between each pair of
hosts is routed using existing Internet routing protocols. To communicate, the client selects
a sequence of relays, constructs a persistent \emph{circuit} through them, and uses it to establish
a connection to the destination. 
The circuit is constructed iteratively and is encrypted once for every relay, which prevents each
relay from learning more than the previous and next hops, and in particular it prevents any one
relay or local network observer from identifying both the source and destination.
Servers can remain anonymous by running as \emph{hidden services},
which maintain persistent circuits into the anonymity network through which they can be contacted.
Onion routing is well-known to be vulnerable to a
\emph{traffic correlation attack}~\cite{onion-routing:pet2000}, however, in which an adversary that
observes both the client and the destination can deanonymize the connection by correlating the
traffic leaving the client with that entering the destination.

Tor~\cite{tor-design} is the most popular system implementing onion routing.
The Tor network currently consists of over \commanum{7000} relays cumulatively
forwarding 100 Gbps of traffic~\cite{tormetrics}. We will apply \sysname attacks to Tor as
well as to several recent proposals to improve Tor's security by changing the way that it selects
paths: DeNASA~\cite{denasa-pets2016}, TAPS~\cite{taps-ndss2017}, and
Counter-RAPTOR~\cite{counterraptor-sp2017}. These attacks are likely to also
be effective for other similar
proposals~\cite{tor-as,jsdm11ccs,lastor,juen-masters,astoria-ndss2016,mator-popets16}.

{\bf Network-Layer Anonymity Protocols.} \label{subsec:nap}
Onion routing protocols are run at the application layer, which allows them to be
deployed without changes to the existing Internet. However, several
protocols~\cite{oakland2012-lap,pets14-dovetail,hornet-ccs2015,phi-popets17} propose operating at
the network layer for efficiency and ubiquity. These protocols change the way that routers set up
and route packets and thus require changes in some of the core infrastructure of the Internet.
Several of them make use of some other next-generation routing algorithm
(\eg{} pathlets~\cite{pathlet-sigcomm2009}) to propagate routing information and select routing
paths. These protocols have many similarities with onion routing, and it is also useful to view them  through the temporal lens. We focus on two of these protocols, HORNET~\cite{hornet-ccs2015} and Dovetail~\cite{pets14-dovetail}, because they represent two distinct approaches that have been suggested. As indicated in Table~\ref{tab:overview}, our results also imply vulnerability in the
other network-layer anonymity protocols.

\vspace*{-.25in}
\section{Models and Metrics}
\label{sec:models}

\begin{table*}[!ht]
\centering
\caption{Overview of \sysname attacks showing the attacked system, adversary capabilities, attack goal, and evaluation metric.}
\begin{tabularx}{\textwidth}{p{0.5cm} | p{1.4cm} p{2.25cm} p{4cm} p{2.7cm} p{4cm}}
 & System & Adversary & Supporting Capabilities & Attack Goal & Metric\\
\hline
\S\ref{subsec:tor} & Vanilla Tor & Single AS & & Observe client \newline directly & Probability of observing \newline client-guard connection\\
\hline
\S\ref{subsec:counter-raptor} & Counter-RAPTOR & Single AS & BGP hijack & Observe client \newline directly & Probability of hijacking \newline client-guard connection\\
\hline
\S\ref{subsec:hornet_mobility} & HORNET & Destination AS & Links client connections\newline Has identified mobility dataset & Link pseudonym to real-world identity & Accuracy and rejection rate when guessing client identity\\
\hline
\S\ref{subsec:denasa_multiple_conn} & DeNASA & Some Tor relays & Links client connections & Identify client AS  & Entropy of posterior \newline distribution over ASes\\
\hline
\S\ref{subsec:dovetail_multiple_conn} & Dovetail & Single AS & Links client connections & Identify client AS  & Set size of possible client ASes\\
\hline
\S\ref{subsec:taps} & TAPS & Destination \newline website & Induces linkable circuits\newline Performs guard discovery & Identify client AS  & Set size of possible client ASes\\
\hline
\S\ref{subsec:hornet_routing} & HORNET & Destination AS & Tracks connection across \newline routing change & Identify client AS  & Set size of possible client ASes\\
\hline
\S\ref{app:mobility-denasa} & DeNASA & Single AS & & Observe client \newline directly & Probability of observing \newline client-guard connection\\
\hline
\S\ref{app:user-counter-raptor} & Counter-RAPTOR & Some Tor relays & Links client connections & Identify client AS  & Entropy of posterior \newline distribution over ASes\\
\hline
\S\ref{app:user-phi} & PHI & Single AS & Links client connections & Identify client AS  & Accuracy and rejection rate when guessing client AS\\
\end{tabularx}
\label{tab:attacks}
\vspace{-.2in}
\end{table*}

\vspace*{-.15in}
The \sysname attacks each exploit some temporal dynamic, but they differ
in where and how it is exploited. An overview of the attack
methods appears in Table~\ref{tab:attacks}.

{\bf Adversary Models.}
Motivated by mass-surveillance concerns, we consider the context 
of a patient adversary that is interested in performing long-term 
deanonymization attacks. We consider several different adversaries, 
depending on the system and the way a temporal dynamic
enables an attack. In each case we identify a fairly weak adversary 
\emph{within the attacked system's
threat model}, with results in the following threat models: a single AS, at least one Tor relay,
and the destination site. Our adversaries are generally passive, with the exceptions that we
consider an active IP prefix hijack (Section~\ref{subsec:counter-raptor}) and that active methods may
be used to link together connections as originating from the same client.

In several of our attacks, we assume that the adversary has this ability to
link together multiple connections (see Table~\ref{tab:attacks}). Observations at the
Tor-protocol level can allow circuits to be linked via
timing~\cite{latencyleak-tissec,huhta-thesis2014}, but the application
layer enables even more effective linking in many important use cases, including (1) a
malicious website, which can either passively observe \emph{pseudonymous} logins or
actively create linkable connections~\cite{taps-ndss2017};
(2) a public IRC channel, which makes pseudonymous activity observable by any
adversary~\cite{onions-target-wpes17};
(3) a hidden service that a malicious client can repeatedly connect
to~\cite{hs-attack06}; and
(4) an administrative service that only one entity has access to, such as SSH access to a
personal server, where such accesses could be observed and recognized by a malicious ISP hosting the
server.

{\bf Anonymity Metrics.}
Due to the diversity of our attacks, we use several types of metrics for their
evaluation (see Table~\ref{tab:attacks} for each attack's metric).

{\it Probability of observing the client connection}:
In some attacks, the adversary attempts to observe the traffic between the client and the anonymity
network. Such observations can facilitate attacks like website fingerprinting~\cite{forest-wf-sec2016} 
and timing analysis~\cite{ccs2013-usersrouted}. We quantify this attack as the probability that 
the adversary succeeds in observing the client connection.

{\it Size of source anonymity set}:
In cases where the adversary uses his observations to reduce the set of \emph{possible} sources,
we measure anonymity as the size of this set~\cite{stop-and-go}. The sources in our attacks are
ASes.

{\it Accuracy and rejection rate when guessing source}:
Some attacks score possible sources heuristically and then guess the highest-scoring source (sources
are ASes in our attacks). Because these methods aren't perfect, the
guess may be wrong. However, we can recognize when the scores are too low to make a confident
guess and reject making one. For this approach (multi-class classification with the reject
option~\cite{pattern-ml-bishop2006}) we measure anonymity using \emph{accuracy} and
\emph{rejection rate}~\cite{nadeem2009accuracy}. Accuracy is the fraction of correct guesses among
cases in which a guess is made, and rejection rate is the fraction of cases in which no guess was
made.

{\it Entropy of source distribution}:
When the adversary's observations allow
him to perform Bayesian inference on the source, we measure anonymity as the
entropy of the posterior distribution~\cite{Serj02}. Our distributions are over ASes, and we use a
uniform prior. Bayesian inference is a strong deanonymization technique, but it is not feasible for
all attacks (\eg{} due to computational constraints), which prevents us from using this metric in
many cases it might otherwise apply.

Note that many of our metrics measure anonymity of the \emph{client AS}.
Although a single AS may serve many thousands of clients, an attack that identifies the
client AS is still quite dangerous, as (1) the client AS can be targeted to
divulge the user's real identity (such targeting would likely be necessary for real-world
deanonymization even if the client's IP address were known); (2) the diversity of user
attributes (\eg{} physical location) is much lower within a client AS and may combine
well with auxiliary knowledge; and (3) the client AS can be used to link connections
and build a pseudonymous profile.
Indeed, a main challenge in designing network-layer anonymity protocols is hiding the client AS,
as the IP address can be easily hidden by the client AS using Network Address Translation.

In several analyses, we focus on the anonymity of the users against whom the attack is \emph{most}
successful, such as those
in locations that experience the largest anonymity losses. The vulnerability of such users to
deanonymization is important to consider, as
(1) users don't benefit equally from anonymity, and the most vulnerable users may
suffer the most from deanonymization; (2) even if a minority of users will end up being
deanonymized, that small risk may be too high for a \emph{majority} of users, who thus
cannot use the system; and (3) relatively few deanonymizations may erode overall trust in the
system given the difficulty of communicating inconsistent anonymity guarantees.

{\bf Network Model.}
To model Internet routing, we generally infer an AS-level topology. We infer AS paths using 
the algorithm proposed by Mao \etal{}~\cite{as-path-inference-sigmetrics05}, which
searches for shortest \emph{valley-free} paths that respect local preference for different
economic relationships (\eg{} customer, provider, peer).
While this type of inference
isn't perfect~\cite{quicksand-ccr2012}, it is used in the original evaluations of all the
recently-proposed anonymity systems that we study (Tor, being an older design, originally
omitted any network-level analysis at all but has since been evaluated using traceroute
data~\cite{tortraceroutes-pets15}). This is true even for those systems (HORNET, Dovetail) that do
not work with BGP, which are evaluated using the existing Internet topology under the supposition
that a future Internet would have similar topological properties.

\vspace*{-.25in}
\section{Client Mobility} \label{sec:mobility}

\newcommand{\IQRRange}{$\left[Q_1 - 1.5\ \IQR{},\ Q_3 + 1.5\ \IQR{}\right]$}

\vspace*{-.15in}
We demonstrate how an adversary can exploit information leakage from client movements to
deanonymize client communications. As clients connect to anonymity networks from various network
locations, they expose themselves to more adversaries and leak location data that will enable
adversaries to deanonymize them. We experimentally quantify the effectiveness of such
attacks for Vanilla Tor, Counter-RAPTOR, and HORNET using
real-world Foursquare and Gowalla data to model client mobility.
We also present supplemental results for DeNASA in Appendix~\ref{app:mobility-denasa}.

\vspace*{-.25in}
\subsection{Mobility Dataset}
\label{subsec:mob_dataset}

\vspace*{-.15in}
\begin{table}[ht!]
\renewcommand{\tabcolsep}{3.5pt}
\begin{center}
\vspace{-3mm}
\caption{Number of users with country-level movements and number of days to
   complete the movements in Foursquare (\textbf{F}) and Gowalla (\textbf{G}) datasets.}
\vspace{-3mm}
\label{tbl:user_country}
\small
   \begin{tabular}{ l l | c c c c c c c c }
      Num. Countries & & 2 & 3 & 4 & 5 & 6 & $\geq$ 7  \\ \hline
      Users & F & 40145 & 13179 & 5649 & 2708 & 1490 & 2574 \\
      Users & G & 17884 & 4557 & 1694 & 705 & 305 & 299 \\
      $Q_1$ Days & F & 48 & 120 & 195 & 228 & 248 & 245 \\
      $Q_1$ Days & G & 7 & 31 & 56 & 77 & 103 & 125 \\
      Med. Days & F & 144 & 252 & 301 & 331 & 353 & 364 \\
      Med. Days & G & 24 & 71 & 111 & 135 & 160 & 177 \\
  \end{tabular}
\end{center}
\vspace{-.2in}
\end{table}

We explored two datasets to model client movements: a Gowalla Dataset \cite{cho2011friendship}  and a Foursquare Dataset \cite{yang2016participatory,yang2015nationtelescope}.
The datasets contain location data from %
real users over the periods Feb~2009--Oct~2010 and Apr~2012--Sep~2013, respectively.
We focus on the country-level movements of users and use it as a proxy for network movements, as we lack fine-grained data to model AS-level movements.

Table \ref{tbl:user_country} compares the two datasets in terms of numbers of users with distinct country-level movements. Note that we only count \emph{new} countries in the number of countries visited, \ie{}, if a user moves out of a country and later travels back, we do not recount it.
The table also shows the Quartile 1 ($Q_1$) and median number of days it takes to visit each number of countries in both datasets.
The median time may take months due to some infrequent travelers, but there are some clients who visit several countries in less than a month.
We also notice that the Foursquare dataset shows a higher $Q_1$/median number of days, indicating that Foursquare users travel less frequently than Gowalla users. However, given the large number of Foursquare users, the absolute number of users for a given travel frequency is similar between the two datasets.

For our analysis in this section, we use both the Foursquare and Gowalla datasets to model client movements.
We use two datasets to map
each country to a ``possible'' AS that a user may connect from in that country:
(1) for evaluations on the Tor network, we use Juen's top Tor client ASes dataset \cite{juen-masters} to map the country to the top Tor client AS located 
in that country, and
(2) for network-layer anonymity protocols, we use the top Internet Service Provider per country (based on the number of active IP addresses) \cite{ipinfo}. 

\vspace*{-.25in}
\subsection{Vanilla Tor} \label{subsec:tor}

\vspace*{-.15in}
{\bf Protocol.} %
A Tor circuit typically consists of three hops. Clients choose a small set of relays
(the default number is one) called \emph{guards} that are used as the first hop in every circuit.
A client will attempt to use the same guard for four to five months before choosing new guards, but this rate
is often accelerated because a client is forced to choose a new guard if its guard goes offline.
To balance the traffic load, Tor relays are
chosen by clients with probability proportional to their bandwidth.

{\bf Attack.}
We quantify the probability that, as a client moves, an adversary is able to observe the client's
Tor traffic from the critical position between the client and its guard. Such a position allows the adversary
to observe the client IP address and thus to perform website
fingerprinting~\cite{forest-wf-sec2016}, to locate hidden services~\cite{hs-attack06}, and to
deanonymize via traffic correlation when destination observations are also
available~\cite{ccs2013-usersrouted}. This position is so sensitive that Tor developed entry guards
specifically to make it difficult to observe. We consider a passive adversary
controlling a single AS whose goal is to be on-path
between a client and its guard relay at some point in time.
As a client moves to new locations while still using the same guard relay chosen at the initial location, 
an AS-level adversary will have increasing probabilities to observe the client-guard connection.
Note that, although we suppose that the client uses the same guard across
movements (\eg{} by using Tor on a laptop), a similar increase in client-guard
traffic exposure occurs if the client uses a different Tor instance at each new
location.

{\bf Methodology.}
We obtain Tor network data from CollecTor
\cite{torcollector}. We use a Tor relay consensus file from 1~Oct~2016 and
retrieve relevant data fields from each relay entry, such as guard flag,
bandwidth, and IP address. We map IP addresses to ASes using Route~Views
prefix-to-AS mappings \cite{routeviews-prefix2as}. We use CAIDA's Internet
topology \cite {caidadata} with inferred AS relationships from Oct~2016.  We
use the same data to model Tor and Internet routing throughout the paper unless
otherwise noted.

We evaluate the CAIDA top 50 ASes~\cite{caidaranking} as potential adversaries which include all
Tier 1 ASes.
We choose the top 50 ASes to evaluate because they are large Internet service providers that
carry a significant amount of network traffic, and thus they are at a good position to observe 
client-guard connections (we did also consider the attack from all ASes and observed a similar
increase in probability). 
\emph{We measure the probability of these ASes to be on-path in the client-guard connection at least once during the client's movements. } 
We assume that the client connects to the Tor network at least once in each country. 
For a given client and for each guard, we use AS path inference to determine 
if an adversary AS is on-path at the client's current and previous locations. 
We then compute the probability by weighting across the guards' bandwidth.

{\bf Results.}
\begin{figure}[t]
   \centering
   \includegraphics[width=\columnwidth]{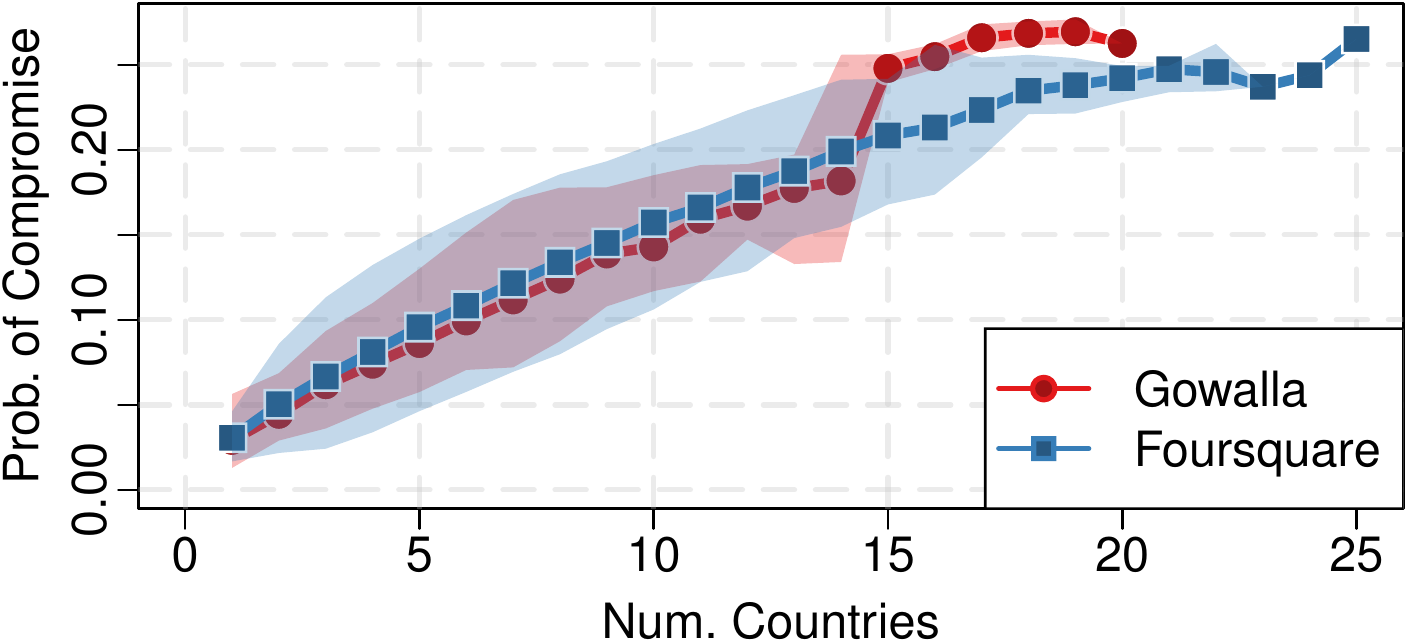}
   \vspace*{-.2in}
   \caption{Probability of compromising a client-guard connection in Vanilla
   Tor, averaged over the top 50 ASes, with the line showing the median and
   shaded area showing values within \IQRRange{}.}
   \vspace*{-.2in}
   \label{fig:vanilla_mobility}
\end{figure}
Figure \ref{fig:vanilla_mobility} shows the probability of compromising a client-guard connection in Vanilla Tor, averaged over the top 50 ASes. %
Each point on the line shows the median attack probability over clients with a given number of
country-level movements. 
The shaded area shows values between \IQRRange{}, where $Q_1$ and $Q_3$ are
Quartile 1 and Quartile 3, respectively, and IQR, the interquartile range, is
defined as $Q_3 - Q_1$. 
This is a standard way to exclude outliers.
Initially, the median probability is 2.8\% with no movement (1 country).
With two more movements (3 countries), the median probability already doubles for both datasets. The probability can reach over 25\% for the clients who have visited 14 countries or more, which is nearly 9 times more than the baseline (1 country, no movement). %
The median probability decreases slightly after 20 country movements in the
Gowalla dataset (23 in Foursquare dataset). This is due to the small sample
of users with high numbers of movements in the datasets, causing higher variance. For
instance,
there is only one Gowalla user that has visited 20 countries, and thus the last data point reflects
the probability of only that user.
Overall, the probability of an adversarial AS compromising a client-guard connection significantly increases during country-level movements.

\vspace*{-.25in}
\subsection{Counter-RAPTOR} \label{subsec:counter-raptor}

\vspace*{-.15in}
{\bf Protocol.} Counter-RAPTOR ~\cite{counterraptor-sp2017} improves Tor's
security against BGP hijacks~\cite{raptor-usenix2015} by changing the way that
guards are chosen. For each guard $G_i$, a client calculates a resilience value
$R_i$ that estimates the fraction of Internet ASes that wouldn't succeed in
hijacking the client's traffic to $G_i$ by (falsely) claiming to be the origin AS
of the IP prefix containing $G_i$. 
The guard relay selection algorithm combines the resilience value $R_i$ and the
bandwidth of the guard $G_i$ by a configurable parameter $\alpha$ in order to take into account both the resilience to hijack attacks and load balancing.

{\bf Attack.} %
We consider an adversary controlling a single AS whose goal is to observe the sensitive client-guard
traffic (as in Section~\ref{subsec:tor}) via a BGP hijack of the guard's prefix.
Counter-RAPTOR aims to maximize Tor clients' resiliencies
to hijack attacks by choosing a guard relay based on client location. However, the guard
selection is done based on the  client's \emph{initial} location, and the same guard is used for
several months \emph{even though the clients may move across locations}.
Thus, AS-level adversaries have an increased power for succeeding in a hijack attack
when clients move to new locations because the guard is only optimized for the \emph{initial}
location.

{\bf Methodology.}
We evaluate the CAIDA top 50 ASes~\cite{caidaranking} as potential adversaries.
We measure the probability of these ASes to succeed in at least one hijack attack during the
client's movements. Note that a successful hijack attack allows the adversary to observe the
traffic between client and guard, enabling the traffic analysis attacks discussed in
Section~\ref{subsec:tor}.
For each client and each guard, we use AS path inference to determine
if an adversary AS can succeed in a hijack attack on the guard.
We then compute the attack probability by weighting across the bandwidth of the guards.

{\bf Results.}
\begin{figure}[t]
   \centering
   \includegraphics[width=\columnwidth]{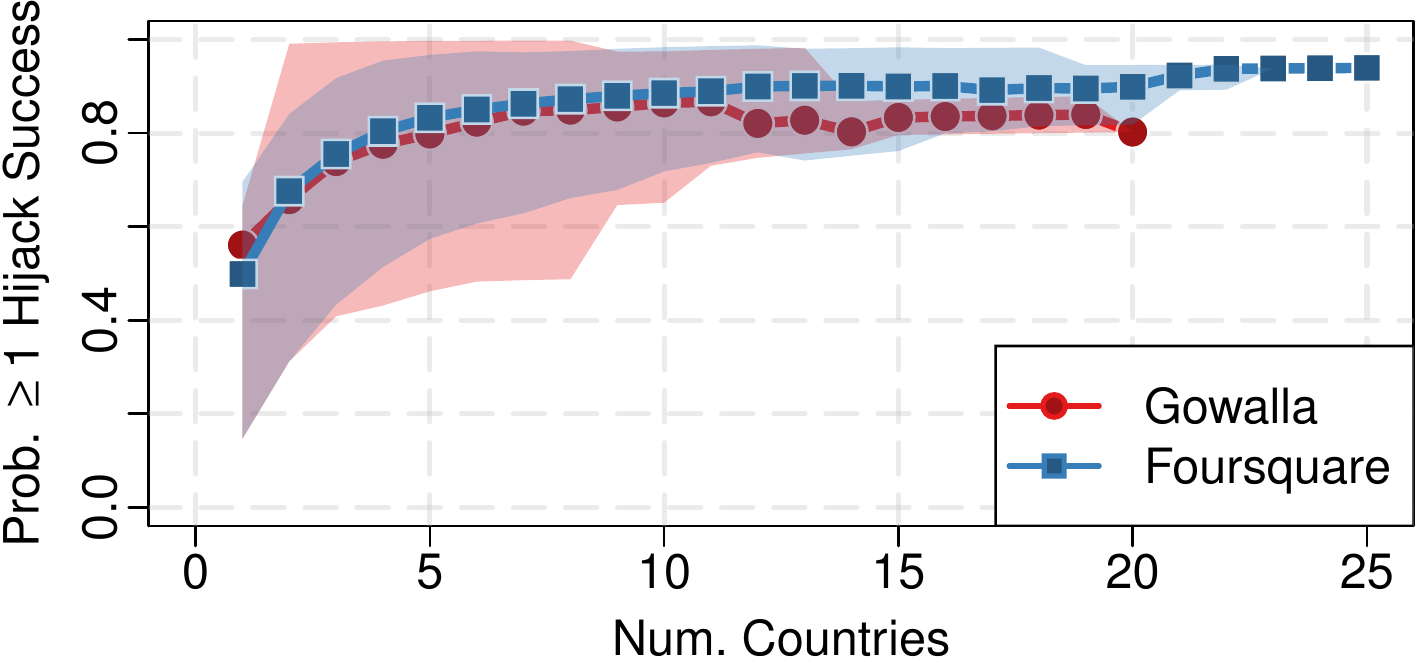}
   \vspace*{-.2in}
   \caption{Probability of succeeding in at least one hijack in Counter-RAPTOR,
   averaged over the top 50 ASes, with the line showing the median and shaded
   area showing values within \IQRRange{}.}
   \vspace*{-.2in}
   \label{fig:cr_mobility}
\end{figure}
Figure \ref{fig:cr_mobility} shows the probability of succeeding at least one hijack attack on the client's guard, averaged over the top 50 ASes.
Each point on the line shows the median value over clients with a given number of country-level movements. 
The shaded area shows values in the range \IQRRange{}.
We can see that the median probability of hijack success reaches about 68\% with only one movement (2 countries) for both datasets, 
compared to an initial of 58\% and 50\%, respectively, with no movement (1 country). 
When the number of movements is more than 10 countries, the attack probability can reach over 90\%. 
The median probability decreases slightly when the number of countries reaches
more than 12 in the Gowalla dataset due to the same reason as in Figure
\ref{fig:vanilla_mobility}; when the number of movements increases, there are
fewer users resulting in higher variance.
Overall, our analysis shows that the probability of hijack success can quickly increase 
with only very few client movements. %

\vspace*{-.25in}
\subsection{HORNET}
\label{subsec:hornet_mobility}

\vspace*{-.15in}
{\bf Protocol.}
HORNET~\cite{hornet-ccs2015} (short for High-speed Onion Routing at the NETwork layer) provides
similar privacy guarantees as onion routing but operates at the network layer. 
HORNET builds on routing protocols in which the source can obtain potential routing paths to the
destination.
Each router cryptographically modifies the full packet (\ie headers and
payload) such that each AS on-path between a source and destination can
identify only its previous and next AS hops.
HORNET's stated threat model is an adversary who compromises a fraction of
ASes, possibly including the destination AS.

{\bf Attack.}
Our attack on HORNET exploits information leakage via auxiliary information about client movements
patterns.  The threat model for this attack is an adversary who compromises the
destination AS.  The adversary wants to deanonymize clients who
use a service hosted within the destination AS, \eg{} an AS who provides cloud
services wants to find the true
identities of pseudonymous accounts on a video streaming website hosted in its cloud.
The clients' identities are protected by HORNET
when logging into their pseudonymous accounts on the destination website.
At the same time, the clients may \emph{reveal} their identities and
locations via auxiliary channels.
For instance, researchers, activists, and politicians who frequently travel to different countries
and give public speeches expose both their identities and locations.
Users may also check in using location-aware services such as Foursquare and Yelp that publicly
reveal that information. Identified user movements may also be collected by a cellular provider and
shared with the website for commercial purposes or with a common legal authority.
Note that the auxiliary location information is not directly linkable to the users' accounts on the
destination website.

The adversary AS has access to two sets of data:
(1) pseudonymous accounts of its clients and the preceding hop in the AS path from the clients to the destination AS;
(2) auxiliary information that contain real identities of people and their location data.

The adversary's goal is to link the two datasets and connect the real identities to the pseudonymous activities of the clients. \emph{The intuition is that the penultimate hop used to reach the destination AS depends on the location of the client.}
With dataset (1), the adversary can identify a set of possible client locations for each connection to a pseudonymous account by considering which locations could choose a path to the destination through the observed penultimate AS hop. Then, with dataset (2), the adversary can examine the location data of identified users and ask the question: \emph{which identities were in one of the possible locations for the connection?} This anonymity set could be quite large if only one pseudonymous connection is considered, but by linking many connections over time the adversary can derive new information when a client moves to new location and thus shrink the client's anonymity set.

{\bf Methodology}. As described in Section \ref{subsec:mob_dataset}, we use the Foursquare and Gowalla datasets as auxiliary information that reveal real client identities and their locations. 
We map each client geo-location to the AS of the top Internet Service Provider that offers service in the 
country of the geo-location. Before any location information, each client's anonymity set comprises the entire 
population of  \commanum{107061} Gowalla users and \commanum{266909} Foursquare users, respectively. 
We then consider location data points at a daily granularity. 
We assume the clients have daily connections to the destination. 
For each day, we take all the clients with location data and compute the penultimate AS hops given their mapped source ASes. Then, for each client, we can eliminate all the other clients without the same penultimate AS hop from its anonymity set. 
Given the imperfect auxiliary information (not all clients have location data everyday), we only eliminate clients with location data the same day, and assume that the clients who do not have data can be at any arbitrary location and thus do not eliminate them from any anonymity set. 

After processing all location data from the dataset, we rank the remaining candidates in each client's anonymity set based on their total number of location data points, from highest to lowest. The intuition is that we want to place more confidence in the clients who reveal their locations frequently and thus provide more identifying information.
We also assign a weight value to each candidate $i$ as $\mathrm{e}^{a * N_i}$,
where $a=0.1$ and $N_i$ is the number of location data points of candidate $i$. The value of $a$ was chosen based on the distribution of number of location data points to scale down the numerical values. Then, for each candidate, we compute the \emph{weight ratio} of its weight over the sum of all candidates' weights in the given client's anonymity set.

Next, we focus only on the highest-weight candidate in each anonymity set. If the weight ratio of the highest-ranked candidate is above a threshold, then we will guess this candidate as the source; otherwise, we ``reject'' the input (\ie{} do not guess) due to a lack of confidence.
Note that we allow a single client to have multiple pseudonymous accounts on the destination website, as we may guess the same candidate for multiple client accounts in the above process.

We evaluate the accuracy rates vs. the number of location data points with fixed rejection rate ~\cite{nadeem2009accuracy} to show the effectiveness of the attack. Accuracy rate is computed as the number of correct guesses over the total number of guesses, and rejection rate is computed as the number of non-guesses over the total number of pseudonymous accounts.

\begin{figure}[t]
   \centering
   \includegraphics[width=\columnwidth]{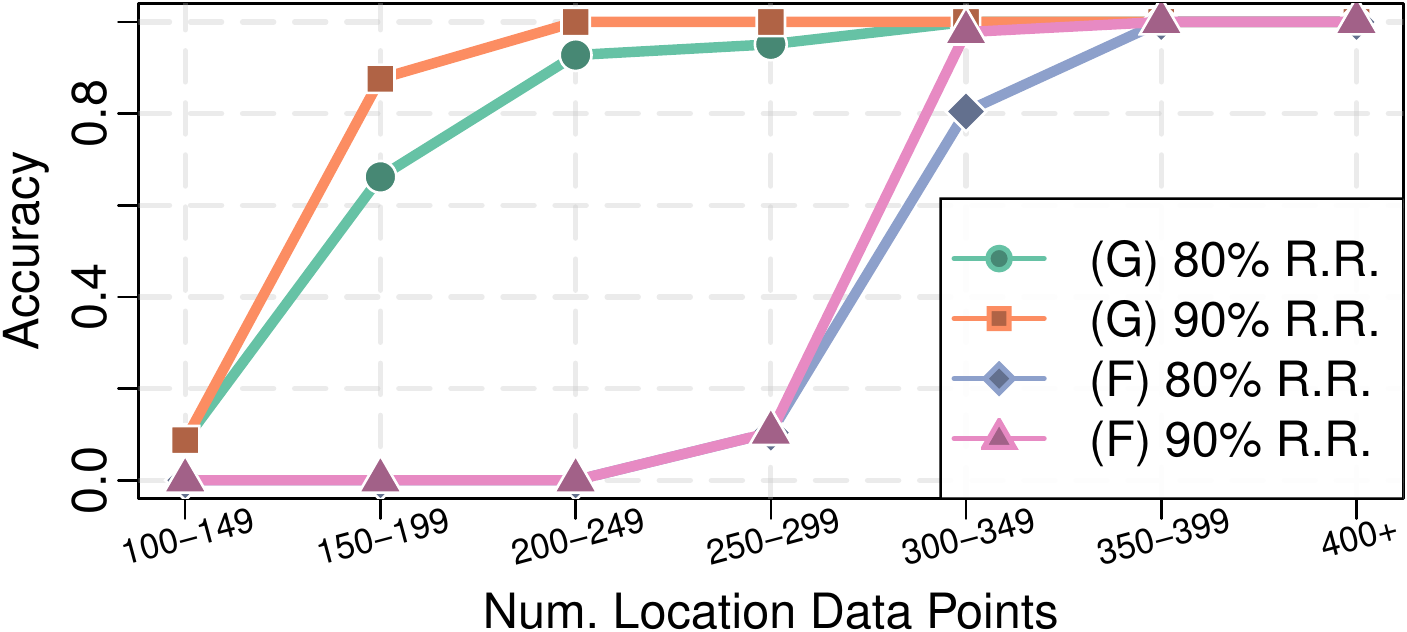}
   \vspace*{-.2in}
   \caption{Accuracy rates for HORNET deanonymizations with increase in number of location data points at various rejection rates (R.R.) for Gowalla (G) and Foursquare (F) datasets.}
   \vspace*{-.2in}
   \label{fig:mob_hornet_twosets}
\end{figure}

{\bf Results.} We present the results for destination Fastly (AS541130), which is a CDN provider for reddit.com and many other websites. 
We also evaluate other popular sites measured by Alexa \cite{alexa-top} such as Google (AS15169), Facebook (AS32934), and Twitter (AS13414), the results for which show similar patterns and appear in Appendix~\ref{app:mobility-hornet}. %
We evaluate clients for which we have sufficient location information,  
which, in this case, are the clients with at least 100 location data points in
the datasets (there are \commanum{50566} such clients in Foursquare and
\commanum{4645} in Gowalla).
Figure~\ref{fig:mob_hornet_twosets} illustrates the results for fixed rejection rates of 80\% and 90\%, respectively, for both datasets. For each data point, we bucket clients using increments of
50 location data points. For clients with more than $400$ location data points, we group them all together since there are many fewer clients.
We can see that the accuracy rate quickly increases with the number of location data points. For Gowalla users, with 200 location data points or more, the accuracy rate reaches 100\% with rejection rate at 90\%, meaning that the adversary can deanonymize 10\% of the clients with no false positives. With lower rejection rate at 80\%, the accuracy rate eventually reaches 100\% as well for clients with 300 location data points or more. 
For Foursquare users, the accuracy rate reaches 97\% at 90\% rejection rate with 300 location data points or more, and reaches 100\% with 350 location data points or more for both rejection rates. 
The difference between the two datasets can be due to the frequency of client movements, e.g., a client with 300 data points of the same location may not reveal that much information compared to a client with 300 data points spread across 10 different countries. From Table \ref{tbl:user_country}, we can see that Foursquare users travel less frequently than Gowalla users.

\vspace*{-.25in}
\subsection{Summary}

\vspace*{-.15in}
We show that client mobility can expose 
new vulnerabilities that put mobile clients at risk. 
For Vanilla Tor, client mobility increases the probability 
that an AS-level adversary can observe the traffic between 
clients and guards; for Counter-RAPTOR, client mobility 
increases the probability of succeeding in routing attacks; 
for HORNET, client mobility makes it very effective and accurate 
to deanonymize clients that reveal a sufficient amount of location information.

\newcommand{\simulationDate}{October 2016}
\newcommand{\numCliqueASes}{\commanum{55244}}
\newcommand{\numLeakyASes}{ten}
\newcommand{\numSamples}{100}
\newcommand{\numDovetailASes}{\commanum{55244}}

\vspace*{-.25in}
\section{User Behavior}\label{sec:multiple_connections}

\vspace*{-.15in}
In some of the anonymity systems that we study, clients leak partial information about
their network location through observable parts of their anonymity paths.
If an adversary learns such information from a
single connection, then an adversary that can \emph{link multiple connections}
as originating from the same client may learn \emph{increasing} amounts of
information.  In this section, we attack two protocols, DeNASA and Dovetail,
with an adversary who links together observations over time.
Additionally, we refer an interested reader to Appendices
\ref{app:user-counter-raptor} and \ref{app:user-phi}, where we provide
supplemental results exploring how our Tempest attacks can be extended to two
similar systems: Counter-RAPTOR and PHI.

\vspace*{-.25in}
\subsection{DeNASA}\label{subsec:denasa_multiple_conn}

\vspace*{-.15in}
{\bf Protocol.}
DeNASA~\cite{denasa-pets2016} is a proposal to improve Tor's security by
modifying how relays are selected for circuits. It is designed to prevent
deanonymization via traffic correlation by a small number of
\emph{Suspect ASes}, that is, ASes that appear frequently on paths to or from
the Tor network. In the DeNASA guard-selection algorithm \emph{g-select},
clients make a bandwidth-weighted choice of guard only from among relays that
are \emph{suspect-free}.  A suspect-free relay is defined as a relay such that
neither of the top two Suspect ASes appear on the network path between the
client and the relay.  Suspect ASes are globally ranked in descending order of
how often they are in a position to perform traffic correlation on Tor
circuits.

{\bf Attack.}
The subset of suspect-free relays available to a client varies depending on
the client ISP's Internet links and routing policies.
Because of this variation across
client locations, clients in different locations will often select the same
guard with differing probabilities.  Thus, a guard selection leaks some
location information; an adversary who can identify a client's guard can
attempt to infer the client's location by considering guard selection
likelihoods from various locations.  In this \sysname{} attack, we demonstrate
that multiple guard selections, identified by the adversary \emph{over time} as
belonging to the same client, can leak enough information to reveal the
client's AS. Recall that identifying the client's AS is a serious degradation of
anonymity (Section~\ref{sec:models}).

To run this attack, the adversary runs some Tor relays.  The malicious
relays will occasionally be used for a client's circuits, allowing the
adversary to discover the client's guards over time
\cite{predecessor-tissec2004}.  The adversary may also employ other known guard
discovery attacks \cite{long-paths, mittal-stealthy}.  We suppose that the
adversary is able to link together a client's connections over time using the
methods described in Section~\ref{sec:models}, which can all be accomplished by
an adversary running relays.  Having identified guards $(G_1,\ldots ,G_n)$ used
by the same client over time, the adversary performs Bayesian inference to
compute the posterior probability $\Pr(L \mid G_1, \dots, G_n)$ for each
possible client location $L$. The adversary uses whatever prior belief he has
about the client location (we use a uniform prior).  Clients use public routing
data to identify suspect-free relays, and so the adversary can compute the
guard-selection distributions for all client locations needed to compute the
posterior distribution over locations.  Increasing numbers of identified guards
reduces the uncertainty in this posterior distribution and can effectively
reveal the client location.

\textbf{Methodology.}
We analyze this information leakage over multiple guard selections by
simulating DeNASA's g-select algorithm using our network model
(Section \ref{sec:models}).  We identify a maximal connected component containing
\numCliqueASes{} ASes in the graph generated from our AS path inference to use
as Tor client locations; we restrict our analysis to a clique to prevent
inaccuracies arising from missing edges in the inferred graph.
For each of these client ASes, we compute the
suspect-free guard selection distribution that clients inside these ASes will
use to choose guards.  Following Barton \etal{} \cite{denasa-pets2016}, we use
AS1299 (Telia Company) and AS3356 (Level 3) as the suspect ASes that clients
try to avoid.

Since the adversary performs Bayesian inference in this attack, we use the
entropy of the adversary's posterior distribution over client locations as our
measure of client anonymity.  We employ a number of heuristics to search for
worst-case ``leaky'' client ASes from which guard selections reveal
location information quickly; for example, one heuristic ranks client ASes in
ascending order by the mean entropy of the adversary's posterior distribution
after a \emph{single} guard selection is made (averaged over guards).  We
select the top \numLeakyASes{} leaky client ASes identified by our heuristics
to evaluate.

\begin{figure}[t]
   \centering
   \includegraphics[width=\columnwidth]{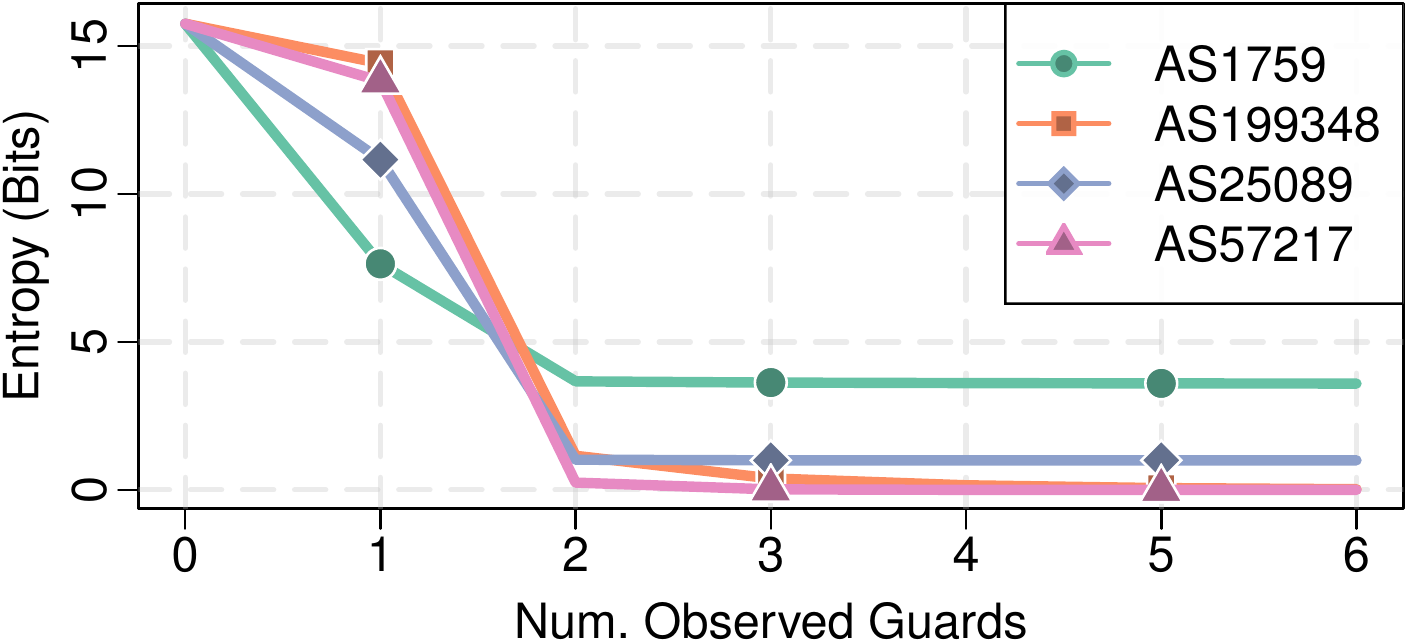}
   \vspace*{-.2in}
   \caption{Mean entropy of posterior client-AS distribution of DeNASA g-select
   clients in ``leaky'' ASes after $x$ guard observations.}
   \vspace*{-.2in}
   \label{fig:denasa_entropy}
\end{figure}

For each of these \numLeakyASes{} client ASes, we simulate a client inside the
AS making guard selections (with replacement).  We compute the likelihood that
these selections were made by a client in each  of the possible
\numCliqueASes{} client ASes.  From these likelihoods, we generate the
adversary's posterior distribution over all possible client locations using
a uniform prior.  We perform this simulation \numSamples{} times for
each of the \numLeakyASes{} client ASes.

\textbf{Results.}
In Figure~\ref{fig:denasa_entropy} we show the entropy of this posterior
location probability distribution, averaged across the samples, for each client
AS at varying number of guard observations.  For visual clarity, we only
present four of the evaluated ASes (the other six ASes exhibit similar or
identical trends).  After two observed guard selections, clients from all ten
ASes drop below 4 bits of entropy.  After three observations, the median
entropy for these ASes drops to just 1 bit.  \emph{Observe how multiple
observations combine to cause dramatic reductions in anonymity.}  For example,
after a single guard selection observation, clients in AS199348 only lose an
average of 1.34 bits of entropy; however, a second observation, in combination
with the first, yields a substantial loss of 13.26 bits and depletes nearly
all of the client's entropy.  A client is forced to select two or more
guards quickly (\eg{} in hours or days) if his guards go offline
and is guaranteed to select a new guard every few months.

\vspace*{-.25in}
\subsection{Dovetail}\label{subsec:dovetail_multiple_conn}

\vspace*{-.15in}
\textbf{Protocol.}
Dovetail~\cite{pets14-dovetail} is a network-layer anonymity protocol
designed for networks that support source-controlled routing.
In Dovetail, a source host $S$ builds a route to a destination host $D$ using
overlapping \emph{head} and \emph{tail} path segments to and from a
\emph{matchmaker} AS $M$, which is chosen randomly from a set of available
matchmaker ASes.  $S$ builds the head path segment to $M$ and encrypts $D$'s
identity to $M$. $S$ and $M$ coordinate to choose and build the tail segment
to $D$ such that head and tail path segments intersect at a common AS, $X$,
called the \emph{dovetail}.  After $X$ removes the loop to the matchmaker, the
final path used for communication is $S \leadsto X \leadsto D$.  ASes on a
Dovetail path learn: (1) the identities of their immediate predecessor and
successor ASes on the path, (2) their absolute position in a path (\eg{} first
hop, second hop, \etc{}), and (3) the host/AS at the end of their path segment
(ASes on the head segment learn $M$'s identity, ASes on the tail segment learn
$D$'s identity, and $X$ learns both $M$ and $D$'s identity).

In source-controlled routing, a client chooses the AS path used for his
connections from a subset of paths made available by his ISP --- we call this
path subset \emph{routable paths}.
To build a Dovetail
connection, a client randomly chooses a head path segment from among his
routable paths containing six or more ASes to the matchmaker; the dovetail is
chosen to be the second-to-last AS on the head path segment.  The details of
tail path selection are unimportant for this attack.

\textbf{Attack.}
To reiterate, a client is limited to select a head path segment to the
matchmaker from among routable paths; the availability of routable paths varies
depending upon the client's ISP, and so path usage inherently leaks location
information.  If the adversary compromises the $k$th AS on a
path, then he can narrow down the client's location by considering the set of
ASes that could have created a path of length $(k - 1)$ to the predecessor AS
$P$ in the path.

With our \sysname{} attack, we demonstrate that this leakage is greatly
exacerbated if a single Dovetail client makes many connections.
Suppose the adversary controls an AS that monitors
connections passing through his AS, a scenario that is within Dovetail's stated
threat model.  We require that the adversary links together connections as
originating from the same client (\eg{} using a method described in Section
\ref{sec:models}). The adversary will use his path observations when he happens
to be chosen in the dovetail position
because this position is the closest to the source that also learns the
connection's destination, which facilitates the connection-linking
required for this attack.  Suppose the adversary makes observations $\{(P_1,
k_1), \ldots, (P_n, k_n)\}$ with respect to a client, where $(P_i, k_i)$
denotes the adversary's predecessor and absolute position on the
$i$th connection on which he is in the dovetail position;
then, the adversary can compute the set of possible client locations as
$\cap_{i=1}^{n} \mathcal{L}(P_i, k_i)$ where $\mathcal{L}$ maps a predecessor
$P$ and position $k$ to the set of client locations that can create a path of
length $(k - 1)$ to $P$.  The output of this attack is a \emph{possibility set}
over client locations, in contrast to the \emph{probability distribution} over
client locations obtained by the adversary in the Section
\ref{subsec:denasa_multiple_conn} attack; computing the observation
likelihoods required for Bayesian inference is computationally expensive given
Dovetail's route selection scheme.

\textbf{Methodology.}
We run our attack in a simulated Dovetail network with paths inferred from
CAIDA's Internet topology.  We follow Sankey and Wright and use a network
model in which a client can route a connection through \emph{any} valley-free
AS path between the source and destination with at most one peer-to-peer
link.  ASes without customer ASes act as possible client ISPs in this
analysis; there are \commanum{47052} such ASes in the topology.

\textit{Dovetail Frequency.}
We assume the adversary compromises a single, fixed AS; as such, interesting
ASes to consider as compromised are ones that are selected often as a dovetail.
We run simulations to determine ASes that are likely
to be selected to serve as the dovetail.  In each simulation, a source AS and
matchmaker AS are chosen uniformly at random from among all client ISP ASes and
all ASes, respectively.  Then, we simulate a client in the source AS who builds
a path to the matchmaker AS and record the dovetail AS.
We collect samples by repeating this procedure \commanum{10000} times and choose
as our adversarial AS the most-common dovetail.

\textit{Anonymity Evaluation.}  We run simulations to measure
the efficacy of this attack with respect to the
fixed adversarial AS.  In each simulation, we choose a source AS uniformly at random
from among all client ISP ASes and choose 500 ASes uniformly at random from
among all ASes to serve as matchmaker ASes.  Then, we simulate a client in the
source AS who makes up to 100 repeated connections to the same destination (the
destination location is irrelevant for this analysis).
For each connection, the client chooses a matchmaker AS uniformly at random
from among the 500 possible and builds a path to the matchmaker.  If the fixed
adversarial AS happens to be placed in the dovetail position on this path, the
adversary observes his predecessor AS in the path and his ordinal position in
the path.  The adversary then computes the set of all ASes who could have
constructed a path of observed length to his predecessor and, through
intersection, updates his set of possible client locations maintained
persistently for the entire simulation (initially containing all
\commanum{47052} client ISP ASes).  If the adversary is not selected as the
dovetail for a connection, he simply takes no action.  We record the
size of the possible client location set after each of the 100
source-matchmaker paths are constructed.  We collect samples by repeating this
procedure \commanum{500} times.

\textbf{Results.} \textit{Dovetail Frequency.}  AS1299 (Telia Company) is
selected most frequently as the dovetail AS, used in 4.03\% of all samples.
The distribution is very right-skewed --- although \commanum{1005} unique ASes
are selected as dovetail at least once, the top ten ASes of the distribution
are used as the dovetail in 30.73\% of all samples.

\begin{figure}[t]
   \centering
   \includegraphics[width=\columnwidth]{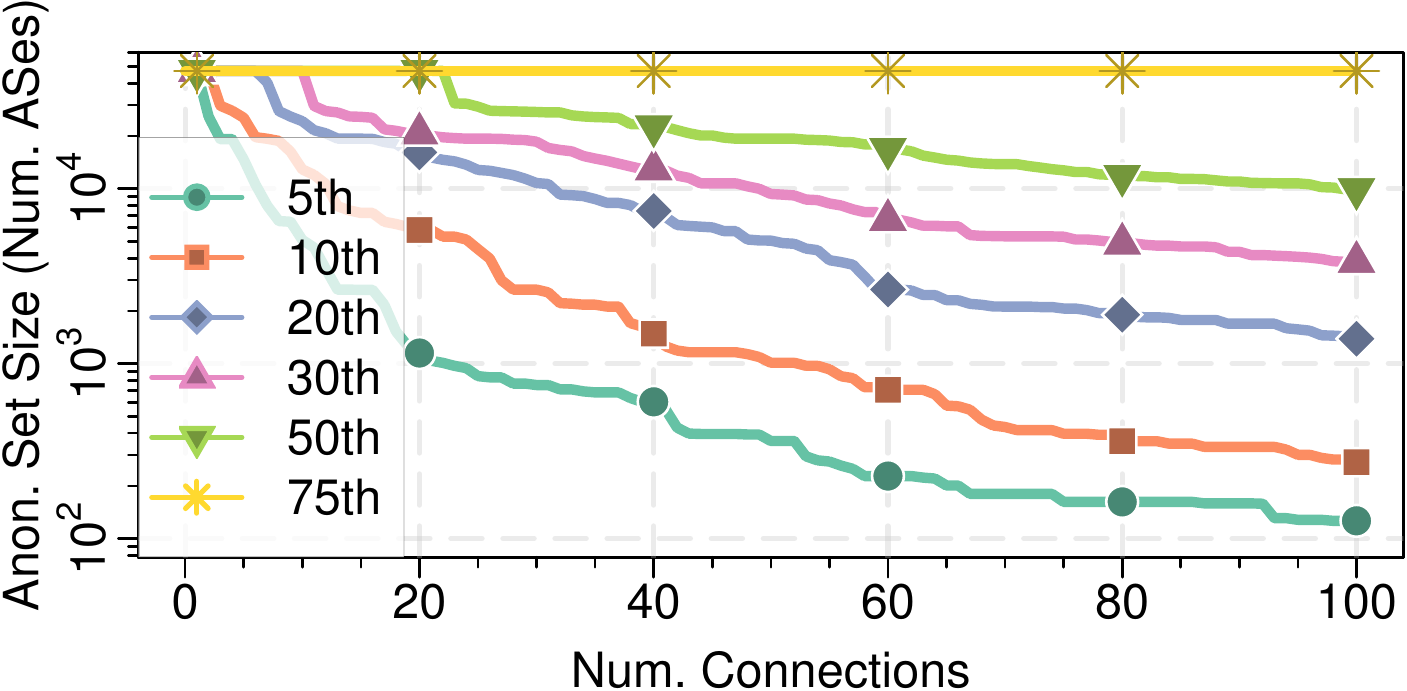}
   \vspace*{-.2in}
   \caption{Anonymity set sizes in Dovetail after $x$ repeated connections with
   respect to AS1299.  Each line corresponds to a percentile of the samples'
   anonymity set sizes.}
   \vspace*{-.2in}
   \label{fig:dovetail}
\end{figure}

\textit{Anonymity Evaluation.}  Because AS1299 was selected as the dovetail most
frequently in our frequency analysis, we use it as our fixed adversarial AS.
We also ran experiments for the other nine most-frequently-selected dovetail
ASes and find that they all produce similar results.  Figure \ref{fig:dovetail}
plots the number of ASes in the source's anonymity set after $x$ repeated
connections.  Each line corresponds to a percentile; for
example, the point at $(x = 100,\ y = 126)$ in the ``5th''
percentile indicates that, after 100 connections, AS1299 could rule out all but
126 possible source ASes in 5\% of samples.  This attack yields significant
anonymity reductions in many of our samples.  After 10 repeated connections, in
5\% of samples, our adversary can rule out all but \commanum{5100} source ASes
as possible --- a 90\% reduction in anonymity set size.  By 100 repeated
connections, the 5th percentile across samples is reduced by
99.7\%.

These results exhibit a continual reduction in anonymity as the client make
connections, suggesting that clients who make many connections place themselves
at a high risk for deanonymization.  A client could make hundreds of repeated
connections over a few days/weeks if he is a frequent user of some Internet service.
We find that each of the ten most frequently selected dovetail ASes are strong
vantage points for this attack.  Many of these ASes are Tier 1 or large transit
networks; such ASes are naturally of interest to many adversaries and are
high value targets for compromise.

\vspace*{-.25in}
\subsection{Summary}

\vspace*{-.15in}
For both the DeNASA and Dovetail protocols, we demonstrate serious weaknesses
to an adversary who can link observations together.  We emphasize that the
significant anonymity degradation in our findings occurs only after a
client makes \emph{multiple connections}, whereas prior work has focused on
quantifying anonymity after a single observed connection.

\newcommand{\rrc}{\texttt{rrc01}}
\newcommand{\numTapsASes}{\commanum{50388}}
\newcommand{\numHornetStableASes}{\commanum{2726}}

\vspace*{-.25in}
\section{Routing Changes} \label{sec:routing_changes}

\vspace*{-.15in}
Natural Internet routing dynamics can change the network paths between
hosts in an anonymity system.  In this section, we show how an adversary can use
such route changes to deanonymize users of two systems:
TAPS and HORNET.

\vspace*{-.25in}
\subsection{TAPS} \label{subsec:taps}

\vspace*{-.15in}
{\bf Protocol.} Trust-Aware Path Selection (TAPS)~\cite{taps-ndss2017} proposes
to improve Tor's resistance to traffic-correlation attacks by choosing circuits
based on the trust that a client has in different network components.
TAPS models a \emph{trust belief} as per-adversary probability
distributions that describe the likelihood that relays and network paths
between relays are under observation by each adversary.  Because different
client locations use different network paths, TAPS leaks
information about the client's location through how the client chooses its
circuits' relays. To limit this leakage, TAPS clusters client locations around
a fixed set of representative locations, and then each client chooses circuits
as if it were in its cluster's representative location.
The clustering routine is informed by a static
snapshot of the state of the network at the time of cluster generation.
This snapshot contains information about the entities in the trust belief,
such as the Tor relays and the ASes. Importantly, it includes inferred
AS-level routing paths.
Because the state of the
network changes over time, Johnson \etal{} state that clusters should be
reformed periodically.

{\bf Attack.} There exists a significant temporal issue with the TAPS
clustering process: a client's circuit-construction behavior may reveal his
cluster, a client's cluster may change \textit{across cluster reformations},
and the adversary can intersect those clusters to gradually reveal the
client's location.
The TAPS AS-to-cluster assignments are public, and so the adversary can use
them to perform this deanonymization.

For our attack, we suppose the adversary employs a variant of the \emph{Chosen
Destination Attack}~\cite{taps-ndss2017} in which the client repeatedly
connects to a malicious website over a longer timescale.  The adversary runs a
number of web servers hosted in different ASes and includes resources from each
server in the website. Each time the client connects to the site, the adversary
uses his servers, in conjunction with a guard discovery attack
\cite{long-paths, mittal-stealthy}, to observe the client's
circuit-construction behavior.  As demonstrated against Astoria, these
observations can identify the client's representative location and thus his
cluster.  After the adversary determines a client's clusters $(C_1, \dots,
C_n)$ across $n$ cluster formations, then the adversary knows the client's AS
is in the set $\bigcap_{i=1}^{n} C_i$.  As cluster composition varies across
reformations, the client's AS can eventually be revealed (Section~\ref{sec:models}
describes how revealing the AS degrades the client's anonymity).

\textbf{Methodology.}
We measure this degradation by running the TAPS clustering routine
on archived Tor and Internet routing data. We implement TAPS
(specifically, the TrustAll configuration with \textsf{The Man} trust policy)
to conservatively evaluate how it is affected by network changes. In particular,
we omit AS organizations, IXPs, and IXP organizations as possible sources of
compromise, and we use a fixed prefix-to-AS mapping at all times.
These choices should make our anonymity analysis conservative, as anonymity degrades
faster the more that the composition of TAPS clusters changes across
cluster reformations.
By implementing TAPS to use fewer data sources and more static data, we expect
that the clusters generated by TAPS vary less in composition across
reformations.

We perform twelve TAPS cluster formations --- one for every month in 2016.  We use
CAIDA's serial-2 AS relationships and CollecTor's Tor data for each respective
month, and a fixed Route Views prefix-to-AS mapping set from Jan~2016 for
each clustering.  Following Johnson \etal~\cite{taps-ndss2017}, we use the top
Tor client ASes identified by Juen \cite{juen-masters} as medoid centers and
configure TAPS to generate 200 client clusters.  We verify that our TAPS
implementation is deterministic on input (\ie{} changes in AS-cluster
assignments are driven by changes in network data, and not by some use
of randomness within the clustering algorithm).

We identify a set of \numTapsASes{} \emph{stable ASes}, that is, those that
exist in all twelve
clusterings (ASes are only clustered if they exist in the CAIDA relationship
data set for that month).  We measure the anonymity-set size of each of these
stable ASes after a variable number $n$ of consecutive cluster formations; namely,
$\abs{\bigcap_{i=1}^n\mathcal{C}_i(\text{AS}) }$ where $\mathcal{C}_i$ maps an
AS to the ASes in its cluster in month $i$.

\begin{figure}[t]
   \centering
   \includegraphics[width=\columnwidth]{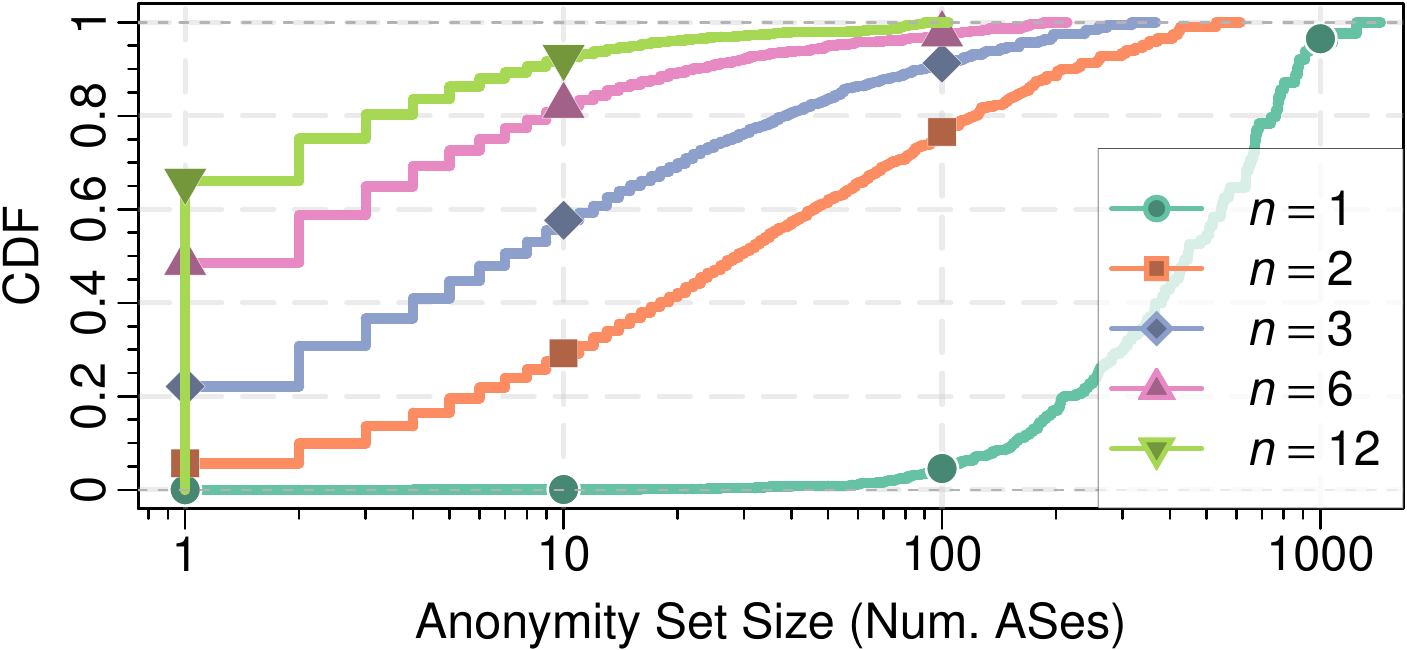}
   \vspace*{-.2in}
   \caption{Anonymity set sizes of possible client ASes in TAPS after $n$
   cluster formations.}
   \vspace*{-.2in}
   \label{fig:taps}
\end{figure}

\textbf{Results.}
Figure \ref{fig:taps} plots distributions of anonymity-set sizes.  Each CDF
corresponds to the distribution after $n$ consecutive cluster formations.  The
$(n=1)$ CDF shows the anonymity-set-size distribution after the initial
clustering in Jan, $(n=2)$ shows the distribution after cluster
formations in Jan and Feb, \etc{} Each point on a CDF
corresponds to the anonymity-set size for one of the \numTapsASes{} client
ASes.

TAPS cluster reformations have serious consequences with respect to user
anonymity.  Initially ($n=1$), the TAPS clustering assignment does well at
producing large clusters with similar sizes: the 5th,
50th, and 95th percentiles are all of the
same order of magnitude with sizes of 103, 445, and 927 ASes, respectively.
However, just a single reformation can harm large fractions of users.  After
just a \emph{single} reformation in Feb ($n=2$), we observe a 94\%
reduction in anonymity-set size, from 445 to 28 ASes, for the median client AS,
and a 99\% reduction in size from 154 to 2 ASes at the
10th percentile.  By six total clusterings, ($n=6$), 49\% of
ASes are left with singleton anonymity sets.  Clients from any one of these
ASes in this near-majority are rendered identifiable to their AS by their
circuit-construction patterns over time.  By being patient and exploiting
this temporal vulnerability in TAPS, the adversary can learn many clients'
locations.

\vspace*{-.25in}
\subsection{HORNET} \label{subsec:hornet_routing}

\vspace*{-.15in}
{\bf Protocol.}
Refer to Section \ref{subsec:hornet_mobility} for a more complete description
of the HORNET protocol.  The important properties of the protocol to
recall are: (1) HORNET provides anonymity at the network layer; (2) the
identities of the source and destination hosts of a connection are hidden from
all ASes carrying the connection's traffic except the
source and destination ASes, respectively; and (3) ASes learn their immediate predecessor and
successor ASes on a connection's path.

{\bf Attack.} For this \sysname{} attack, we observe that network routes can change
over time and that such changes can leak information about the client location.
Consider a compromised AS $A$ that
observes a connection to a destination server within $A$. The adversary
observes the penultimate AS hop $X$ of this connection. Thus, the adversary
knows that the client's location is in the set $S_0$ containing
all ASes that currently have a route to $A$ with penultimate hop $X$.
Now, suppose a route change occurs that causes the client to reconnect through
penultimate AS hop $Y$. We assume that the adversary can link the reconnection to
the prior connection via traffic analysis (\eg{} the destination typically has
only one active connection or a higher-level protocol has an identifiable handshake
pattern at connection start). The adversary can then compute the set $S_1$ containing
ASes that route to $A$ through penultimate hop $Y$ and conclude that the
client's AS is within $S_0 \cap S_1$. Thus, as with the previous attack
(Section~\ref{subsec:taps}), the client's anonymity is degraded by learning its
AS.

To run this attack, the adversary (1) controls an AS $A$, (2)
monitors connections to a third-party destination host within $A$,
and (3) waits for route changes to occur that affect the connections he is
monitoring. The adversary must also have some source of data about the available
routes to $A$, several of which are currently available, including
public routing datasets~\cite{routeviews} and public platforms for traceroute
measurements~\cite{ripe-atlas}.  We assess the risk this attack poses by quantifying the
\emph{frequency} of naturally-occurring route changes and the \emph{impact}
that a single change can have on anonymity.

\textbf{Methodology.}
We use traceroute data made available by RIPE Atlas \cite{ripe-atlas} to
analyze Internet routing changes.  RIPE Atlas is an Internet measurement platform
consisting of thousands of volunteer-run network probes.  These probes are
distributed across thousands of different networks and can be configured to run
various Internet measurements, such as pings and traceroutes.

We consider one such Internet measurement (\texttt{id} $= 5001$); in this
measurement, all online probes run UDP traceroutes to k.root-servers.net.  In
Jan--Feb 2016, the period we consider for this study, this name
resolved to 193.0.14.129 in AS25152.  There were approximately \commanum{8500}
probes hosted across \commanum{5700} IP prefixes, representative of
10--11\% of allocatable IPv4 space, running traceroutes to this destination
every 30 minutes.  For the sake of this analysis, we consider the scenario
where AS25152 is under adversarial control and contains a destination host of
interest to the adversary.

We identify a set of \emph{stable probes} whose IP prefix and AS of origin
do not change during our measurement period; we want to ensure that route
changes that we observe are due to changes in network routes and not
``artificially'' induced by a probe's physical location changing.  We identify
\commanum{6566} stable probes hosted across \numHornetStableASes{} unique ASes.
These probes will serve as the clients our adversary will attempt to
deanonymize.

\textit{Frequency of Changes.}
We use Route Views prefix-to-AS mappings to compute the AS path each probe uses
to reach AS25152.  At the granularity of the measurement interval (30 minutes),
we search for route changes that cause a change in a stable probe's penultimate
hop to AS25152 between two traceroutes.  We perform this search for all stable
probes over the entire month-long period.  We compute the mean number of
penultimate hop changes each AS experiences from Jan--Feb (averaging
over probes).

\textit{Impact of Changes.}
For each route change we identify in a stable probe $p$'s traceroute data, we
compute two sets: $S_0$, containing all probes with traceroutes matching $p$'s
penultimate hop before the route change occurred, and $S_1$, containing
all probes with traceroutes $p$'s penultimate hop after the route change
occurred.  We then measure $p$'s anonymity set size with respect to $S_{0}$ and
$S_{0} \cap S_{1}$ (\ie{} before and after the route change) by computing
$\abs{ \{ \mathcal{A}(p) \mid p \in S_{0} \} } $
and
$\abs{ \{ \mathcal{A}(p) \mid p \in S_{0} \cap S_{1} \} } $,
where $\mathcal{A}$ maps a probe to its Autonomous System of origin.  Following
this method, we compute the mean before-and-after anonymity set sizes for each
AS with at least one route change by averaging over all route changes and
stable probes.

\begin{figure}[t]
   \centering
   \includegraphics[width=\columnwidth]{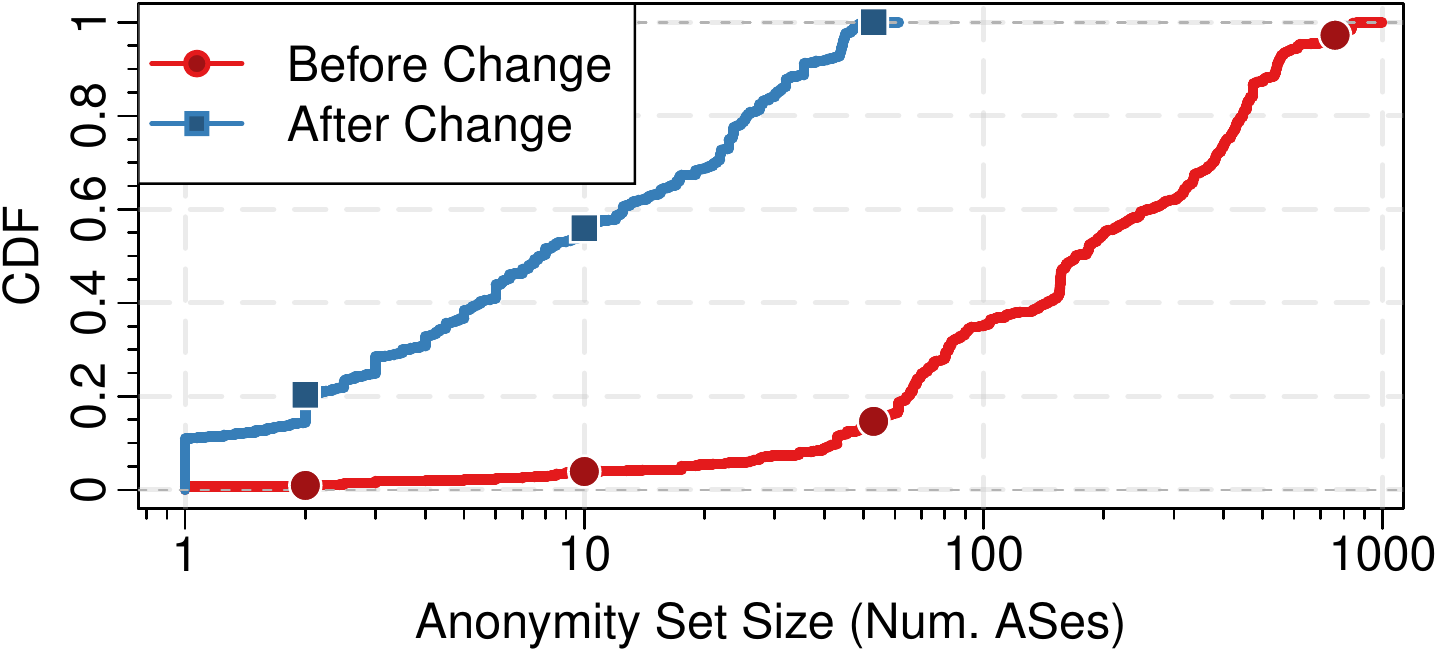}
   \vspace*{-.2in}
   \caption{Mean anonymity set sizes for ASes before and after a route change
   occurs in HORNET.}
   \vspace*{-.2in}
   \label{fig:hornet_churn}
\end{figure}

\textbf{Results.}
\textit{Frequency of Changes.}
We find that 840 (30\%) of the \numHornetStableASes{} ASes experience at least
one penultimate hop route change from Jan--Feb and that the
distribution of changes is long-tailed.  20\% of ASes experience at least 2
penultimate hop changes on average, 10\% of ASes experience at least 3.8
changes, 5\% of ASes experience at least 7.9 changes, and 3\% of ASes
experiences at least 50.8 changes.  ASes that had at least one route change are
susceptible to this attack.  Frequent route changes can further expose users
both by increasing the probability that an adversary can make at least one route
change observation for a user and by introducing the possibility that multiple
route changes can be linked to a single user. Our results are limited to
locations with probes, but, due to the large sample-size of probes,
the distribution of frequencies we obtain should apply to the Internet as a
whole.

\textit{Impact of Changes.}
Figure \ref{fig:hornet_churn} depicts the distribution of mean anonymity set
sizes for each of the 840 ASes with at least one change before and after a
single, average routing change occurs.  A nontrivial percentage of these ASes
(8\%) are left with singleton anonymity sets after the average route change.
There is a sizable reduction in anonymity for most ASes; for example, the
median shifts from an anonymity set size of 170 ASes to just 8 ASes.
The anonymity sets formed in this analysis are limited to
locations with probe coverage.  This particularly affects the absolute
anonymity-set sizes. However, we expect that the \emph{relative reductions}
in anonymity (\eg{} an order of magnitude) reflect an adversary's ability in
practice.

These results present a serious risk for users.  Suppose that the destination
under surveillance in AS25152 serves $c$ live client connections on average at
any given time and suppose that users are distributed roughly uniformly across
the Internet AS space.  From our analysis above, we would expect that $0.3c$
connections will experience a penultimate hop route over the month (see
\textit{Frequency} results, we found that 30\% of source ASes experienced a
penult. hop change).  If the destination is popular or connections are
long-lived, $c$ will be large and thus many users will be vulnerable to this
attack.

\vspace*{-.25in}
\subsection{Summary}

\vspace*{-.15in}
For both TAPS and HORNET, we show that a patient adversary who waits for
network changes (cluster reformations in TAPS, route changes in HORNET) can use
public datasets (cluster assignments, traceroute data) to achieve
order-of-magnitude reductions in clients' anonymity set sizes.  As the network
continues to change, clients ASes can be completely identified.

\vspace*{-.25in}
\section{Discussion and Ethics}

\vspace*{-.15in}
{\bf Countermeasure Challenges.} Explicitly accounting for temporal dynamics could reduce the severity of 
the \sysname attacks. \emph{However, the most straightforward defenses encounter
subtle tradeoffs and weaknesses.} For example, an obvious approach to defending against the
client-mobility attacks on vanilla Tor and Counter-RAPTOR is for the client to select multiple
guards in different locations and use the one closest to current location. However, this raises an
immediate tradeoff between defending against \sysname and limiting the number of potentially
malicious relays in the guard position. This approach also raises the possibility of an attack in
which an adversary places guards in targeted locations to affect nearby clients. As another example,
a natural attempt to prevent the user-behavior attack on
Dovetail would be for each client to choose a small number of matchmakers to use for all
connections. However, as noted by Sankey and Wright (Section 5.1 \cite{pets14-dovetail}), reusing
the same matchmaker gives it the ability to perform an intersection attack across connections
based on their tail segments, as the client chooses each tail segment to minimize intersections with
the head path. As a final example, a plausible defense against the routing-change attack on HORNET
might seem to be for the client to choose paths for which the penultimate hop changes infrequently.
However, in addition to making a client's anonymity dependent on routing dynamics outside of its
control, this creates an additional information leak to the adversary, who could take into account
path variability when considering which client location is a likely source for an observed
connection. Thus defending against the \sysname attacks appears to be a non-trivial challenge that
we leave for future work.

{\bf Active Attacks.} In this paper, we mainly consider adversaries
passively observing network traffic. However,
we do consider an adversary performing active BGP hijacks against Counter-RAPTOR
in Section \ref{subsec:counter-raptor} and Appendix \ref{app:user-counter-raptor},
and we do include active methods among those that might be used to link connections and
perform guard discovery. We leave for future work a more general analysis of the
power of active adversaries to exploit temporal dynamics. Such adversaries could
be very powerful. For instance, an active adversary may
be able to track the movements of a Vanilla Tor client by continually intercepting traffic to
the guard. As another example, an active adversary may cause observable routing changes
by withdrawing and inserting (possibly completely legitimate) routes in HORNET.

{\bf Ethical Considerations.} All the datasets we used in this paper were
publicly available. With the privacy and safety of Tor users in mind, we
refrain from collecting any user data on the live Tor network. Instead, we work
with existing public datasets, such as network data from CAIDA and RIPE Atlas,
Tor data from CollecTor, and location data from Gowalla and Foursquare, to
perform attack analysis while preserving the anonymity of real Tor users.  The
code used for this paper is available at
\texttt{\url{https://github.com/rwails/tempest}}
for review.

\vspace*{-.25in}
\section{Related Work} \label{sec:related}

\vspace*{-.15in}
{\bf Temporal Dynamics.} Similar to our work, there is a thread of research 
that deals with the degradation of anonymity over a period of time. 
In the predecessor attack~\cite{crowds:tissec,predecessor-tissec2004},
an attacker tracks users' communications over multiple path reformulations
and identifies the observed previous hops as the most likely sources of the
connection. {\O}verlier and Syverson~\cite{hs-attack06} used similar 
observations for demonstrating practical attacks against hidden services, 
and motivated the use of and research on guard relays in anonymity 
networks~\cite{cogs-wpes2012,hayes-popets15}. 
Intersection attacks~\cite{raymond:2001,web-mix-pet2000} and disclosure
attacks~\cite{kedogan:ih02,danezis:ih04,mathewson:pet04, danezis:pet09}
aim to compromise client anonymity by intersecting over time the sets of clients
that were active when a given client is observed to receive a message. %
Danezis and Troncoso highlighted the impact of evolution of user behavior in 
disclosure attacks~\cite{danezis:wpes13}.

In contrast to previous works, we identify and analyze novel traffic analysis 
attacks based on exploiting temporal changes in anonymity paths, specifically 
in the context of low-latency anonymity systems and AS-level adversaries. 
We note that our results on \emph{client mobility} represent the first analysis of 
this issue in anonymity systems. Furthermore, our results on exploiting user 
behavior and routing updates represent the first analysis of how \emph{probabilistic 
information leaks due to restricted AS-level Internet topology} can be aggregated 
\emph{over time}. Our work is also unique in systematically investigating these issues 
across a broad range of systems.

{\bf Network-Level Adversaries.} Security analyses of anonymity systems 
typically focus on the threat of end-to-end timing correlation by 
compromised or malicious relays/proxies~\cite{onion-routing-ih96}. Feamster and Dingledine were 
the first to consider this threat from the perspective of network-level 
adversaries~\cite{feamster:wpes2004}. Murdoch and Zieli{\'n}ski~\cite{murdoch:pet2007}
showed that even Internet-exchange 
level adversaries can perform traffic analysis of anonymity systems. Edman 
and Syverson~\cite{tor-as} measured the impact of Tor's path selection strategies on 
security against network-level adversaries. 
Johnson \etal{}~\cite{ccs2013-usersrouted} performed a security analysis of Tor
and measured the risk of deanonymization against both relay-level and 
network-level adversaries over time. Sun \etal{}~\cite{raptor-usenix2015} were the first to 
observe that an adversary could manipulate routing dynamics (BGP) to compromise 
user anonymity in Tor, including exploiting inherent churn in BGP. 
These works have motivated the design of several of the systems we study in this paper,
and our work shows that that the deanonymization risk in those systems is much greater than
previously thought.

{\bf Latency Attacks.}
Hopper \etal{}~\cite{latencyleak-tissec} demonstrated that a malicious
destination can infer a client's location after a number of repeated
connections using information leaked via connection latency in Tor.  Latency
attacks are orthogonal to \sysname{} attacks and can be used in parallel to
enhance the adversary's ability to deanonymize users.  However, \sysname{}
attacks are applicable in cases where latency attacks may be ineffective, \eg{}
in source-controlled routing networks where the availability of many routable
Internet paths may limit information leaks from latency.

{\bf Other Traffic Analysis Attacks.} Our work highlights the risk of abstracting 
away important system components that impact user anonymity in practice. Similarly, 
prior work on traffic analysis has considered a range of related oversights.
Mittal \etal{}~\cite{mittal-stealthy} analyzed the impact of network throughput information, 
and showed that it allows an adversary to infer the identities of Tor relays 
in a circuit. Murdoch and Danezis~\cite{torta05} and Evans \etal{}~\cite{long-paths} considered 
the impact of network congestion on anonymity systems such as Tor. Borisov
\etal{}~\cite{borisov:ccs07} 
and Jansen \etal{}~\cite{sniper14} explored the use of denial of service attacks 
to compromise client anonymity. Murdoch~\cite{murdoch:ccs06} and Zander \etal{}~\cite{zander:usenix08} have shown 
that clock skew can be used for deanonymization.

\vspace*{-.25in}
\section{Conclusion}

\vspace*{-.15in}
We identify temporal dynamics in anonymity paths as potentially degrading the security 
of anonymous-communication systems. We present the \sysname attacks, which make novel 
use of such dynamics in three broad categories, including path changes due to client 
mobility, user behavior, and network routing updates. These attacks are shown to be effective 
against a variety of anonymity systems including both onion-routing and network-layer protocols. 
Our work leads to the following recommendations for the research community: 

{\bf Adversarial Model:} Anonymity systems should consider the threat of a \emph{patient} 
adversary that is interested in performing long-term attacks on anonymity systems. Such an adversary 
can record information about user communications over a long period of time, and then
(1) aggregate probabilistic information leaks over time to deanonymize users, and (2) 
correlate information leaks with auxiliary sources of information, such as data
about client mobility patterns or network routing updates, to deanonymize users. 

{\bf Temporal Dynamics:} Anonymity systems should be analyzed for use over time. 
In particular, system designers ought to consider the effects client mobility,
user behavior, and network routing changes. 
More generally, our work motivates (1) the design of anonymity protocols that are robust in the
presence of temporal  dynamics, and (2) the formalization of security definitions and frameworks
that incorporate relevant temporal issues.

\newpage
\section*{Acknowledgments}
This work was supported by the Office of Naval Research (ONR) and also by the
National Science Foundation (NSF) under under grant numbers CNS-1423139,
CNS-1527401 and CNS-1704105.  The views expressed in this work are strictly
those of the authors and do not necessarily reflect the official policy or
position of ONR or NSF.

\bibliographystyle{plain}
\bibliography{references}

\appendix
\vspace*{-.25in}
\section{Client Mobility} \label{app:mobility}
\vspace*{-.15in}

\subsection{DeNASA} \label{app:mobility-denasa}
\vspace*{-.15in}

{\bf Protocol.}
Refer to Section~\ref{subsec:denasa_multiple_conn}'s protocol description;
briefly restated, DeNASA clients who use g-select choose guards only from among
relays that are suspect-free; \ie{} clients ensure two suspect ASes do not
exist on their client-guard network paths.

{\bf Attack.}
In the similar manner as the attack we outline in Section~\ref{subsec:tor}, we
consider the risk of traffic-correlation by one of the suspect ASes as clients
move between network locations.  Although the client initially chooses a
suspect-free guard, the client may \emph{introduce} one of the suspect ASes
onto his client-guard network path as he moves to new locations while
continuing to use the same guard.

{\bf Methodology.}
We quantify the increasing probability that the suspect ASes will be able to
observe client-guard connections during clients' movements.  We simulate mobile
clients following our mobility model in Section \ref{sec:models}.  Using guard
weights, we compute the probability that a client will choose a guard such that
the client-guard link is compromised by a suspect AS at least once over the
client's movements.  We compute this compromise probability for each client in
both the Foursquare and Gowalla dataset.

\begin{figure}[t]
   \centering
   \includegraphics[width=\columnwidth]{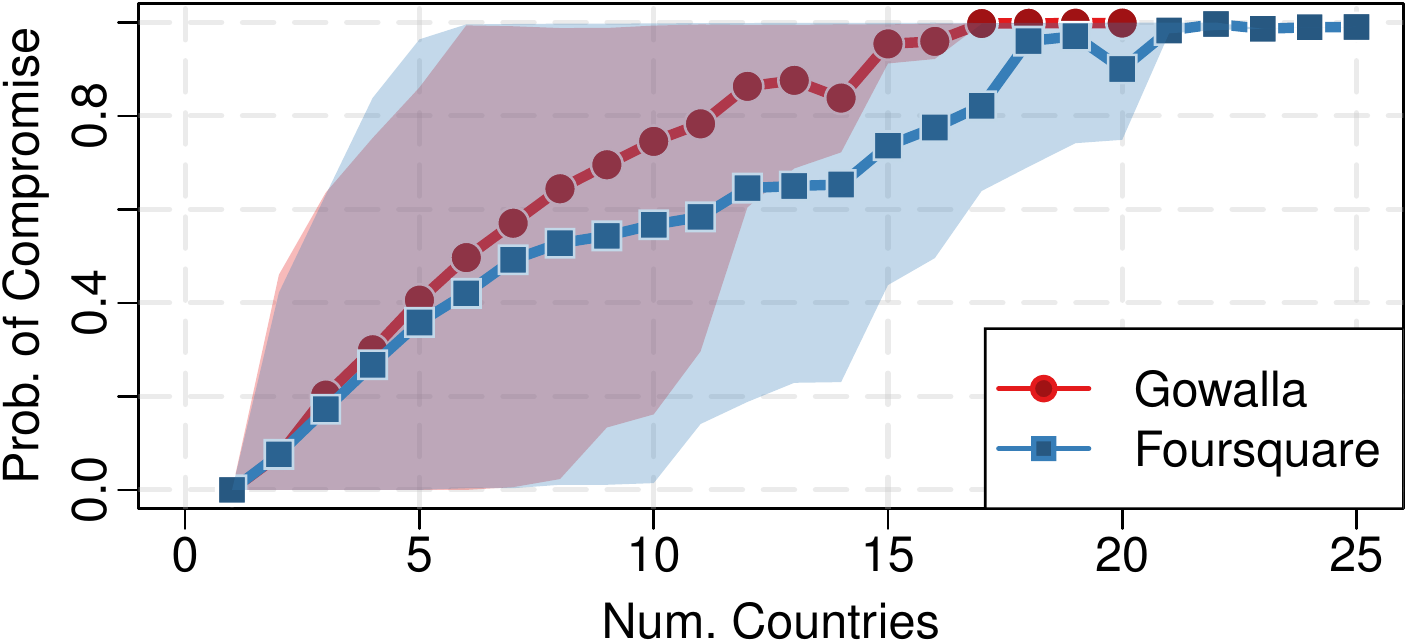}
   \vspace*{-.2in}
   \caption{Probability of compromising client-guard connections for suspect
   ASes in DeNASA.  The line shows the median probability and the shaded area
   shows values within \IQRRange{}.}
   \label{fig:denasa_mobility}
\end{figure}

{\bf Result.}
Figure \ref{fig:denasa_mobility} shows the distribution of clients' probability
of compromise, where clients are grouped together by their number of
country-level movements.  
Each point on the line shows the median probability of compromise over clients with a given number of
country-level movements. 
The shaded area shows values between \IQRRange{}, where $Q_1$ and $Q_3$ are
Quartile 1 and Quartile 3, respectively, and IQR, the interquartile range, is
defined as $Q_3 - Q_1$. 
Initially, all clients are able to
find suspect-free guards, and so clients with no movements have no probability
of being compromised.  Once the clients start moving to different countries, the
probability increases.  There is large variance in probabilities among clients
when the number of country-level movements is relatively small (8 or fewer countries 
for Gowalla users and 10 or fewer countries for Foursquare users) --- for some
clients, the probability remains close to 0, while for others it can get close
to 1 after visiting only six countries in both datasets.  The rate of increase in probability
depends on the countries that a client visits and the order of the visits.

\vspace*{-.25in}
\subsection{HORNET Supplemental} \label{app:mobility-hornet}
\vspace*{-.15in}

\begin{figure}[t]
   \centering
   \includegraphics[width=\columnwidth]{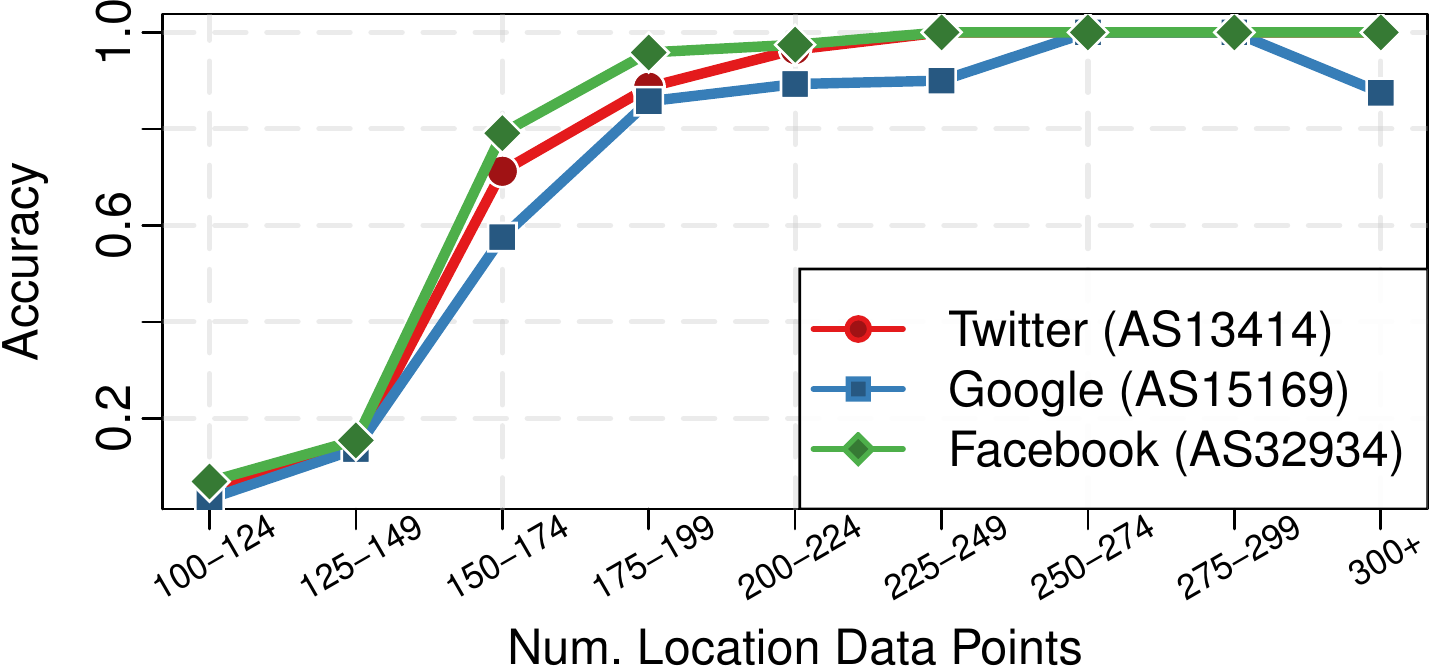}
   \vspace*{-.2in}
   \caption{Accuracy rates for HORNET deanonymization with 90\% rejection rate for Google, Facebook and Twitter in Gowalla dataset.}
   \vspace*{-.2in}
   \label{fig:mob_hornet_appendix}
\end{figure}

In Section \ref{subsec:hornet_mobility}, we present the accuracy rates for
destination Fastly (AS54113) with 80\% and 90\% rejection rates. We have also
performed the same evaluations with respect to Google (AS15169), Facebook
(AS32934) and Twitter (AS13414). We show the results (with 90\% rejection rate)
in Figure~\ref{fig:mob_hornet_appendix} based on the Gowalla dataset.

We can see that the overall trends are similar for all four ASes --- the accuracy rate quickly increases with the number of location data points. 
For both Facebook and Twitter, the accuracy rates reach 100\% when there are at least 225 location data points. 
For Google, the accuracy rate reaches 100\% when number of location data points reaches 250, however, it goes back down to 87.5\% when considering only clients with 300 location data points or more. This could be due to the limited number of clients in the 300-data-point group which leads to lower accuracy rate.

\newcommand{\numPhiASes}{\commanum{55244}}
\vspace*{-.25in}
\section{User Behavior} \label{app:user}
\vspace*{-.15in}

\subsection{Counter-RAPTOR} \label{app:user-counter-raptor}
\vspace*{-.15in}

\textbf{Protocol.}
Recall from Section \ref{subsec:counter-raptor} that Counter-RAPTOR is
another proposed client-location-aware modification to Tor's guard selection.
In Counter-RAPTOR, clients incorporate a BGP hijack resilience value into guard
selection probabilities.

\textbf{Attack.}
Guard resiliency varies across client locations; a guard may be resilient to
hijacks with respect to some client locations but not others.
As such, in the same manner as DeNASA, guard selections that can be
linked to a client by the adversary can be used to infer the client's location.
Counter-RAPTOR does offer some defense against this attack; to prevent relay
load from becoming too skewed by resilience values and to limit location
information leakage, Counter-RAPTOR clients weight guards by a configurable
linear blend of resilience and bandwidth weight.  We run the same \sysname{}
attack described in Section \ref{subsec:denasa_multiple_conn} to evaluate the
efficacy of the Section \ref{subsec:denasa_multiple_conn} \sysname{} attack on
Counter-RAPTOR.

\textbf{Methodology.}
We implement Counter-RAPTOR using archived Tor and Internet topology data.  We
adhere to the adversary model and methodology laid out in Section
\ref{subsec:denasa_multiple_conn}, with a caveat that we use older Tor data
from Oct 2015 for this analysis, but AS topology data derived from CAIDA's
Oct 2016 datasets.  We do not expect that this inconsistency in methodology
significantly affects our results.  We configure Counter-RAPTOR with
configuration parameter $\alpha = 0.5$, \ie{}, clients weight guards 50\% by
their resilience value and 50\% by their bandwidth.  $\alpha = 0.5$ is the
default value recommended by Sun \etal{} \cite{counterraptor-sp2017}.  We
compute the Counter-RAPTOR guard selection distributions for all clients
locations in a fully-connected component of \commanum{55243} ASes.

We perform a search for leaky client locations among the \commanum{55243} ASes
using the heuristic search techniques described in Section
\ref{subsec:denasa_multiple_conn}.  In total, we select the top 18 ASes from
our search heuristics for evaluation.  We perform 100 simulations from each of
these 18 locations of a client selecting up to 500 guards.  In each simulation,
we form a posterior distribution over client locations after each guard
selection, assuming a uniform prior.

\begin{figure}[t]
\centering
   \includegraphics[width=\columnwidth]{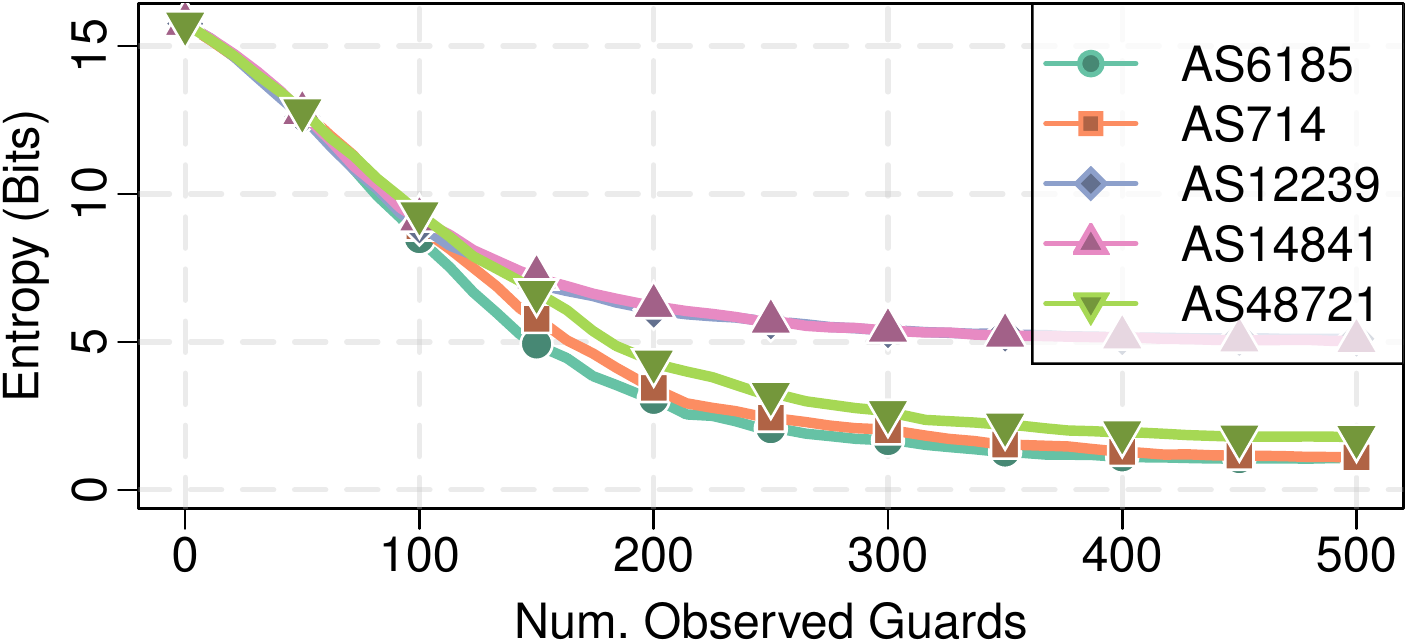}
   \vspace*{-.2in}
   \caption{Mean entropy of posterior client-AS distribution of
   Counter-RAPTOR clients in ``leaky'' ASes over multiple guard
   observations.}
   \vspace*{-.2in}
   \label{fig:counter_raptor}
\end{figure}

\textbf{Results.}
Figure~\ref{fig:counter_raptor} shows the entropy of this posterior client
location probability distribution, averaged across samples, for each client AS
at varying number of guard observations.  For visual clarity, we only include
the five ASes whose samples had the smallest mean entropy after 100 guard
selections.  Guard selections from the leakiest ASes we evaluated for
Counter-RAPTOR are not nearly as informative as we found in DeNASA.  After 100
guard selections, clients in AS6185 are the worst off, but only have lost
$7.44$ bits of entropy on average.  Clients do leak significant amounts of
information over hundreds of guard selections, \eg{} clients in AS174 have 1.1
bits of entropy on average after 500 guard selections; however, it is very
unlikely that a client will need to select this many guards and so the slow
leakage may be acceptable for most users.  Though our search methods are not
exhaustive, these results suggest that Counter-RAPTOR's bandwidth blending may
be a suitable method for limiting location information leaks in location-aware
relay selection algorithms.

\vspace*{-.25in}
\subsection{PHI} \label{app:user-phi}
\vspace*{-.15in}

{\bf Protocol.}
PHI, short for \textbf{P}ath-\textbf{Hi}dden Lightweight Anonymity Protocol, is
a network-layer anonymity protocol that protects users by using encrypted
routing state stored in packet headers and by using path-setup traffic
indirection in a similar fashion to Dovetail.  PHI's design improves upon two
of Dovetail's shortcomings: (1) PHI uses fixed-size packet headers and
randomized placement of routing state within packet headers to prevent ASes
from learning their absolute positions on a path, and (2) PHI offers
compatibility with BGP networks and does not require an infrastructure that
supports source-controlled routing.

In PHI, \textit{helper} ASes serve an analogous role to Dovetail's matchmaker
ASes.  A source host $S$ builds a connection to a destination $D$ using two
\emph{half-paths} to and from a helper AS $H$ which is chosen from a set of
available helpers.  The process of setting up the first half-path from $S$ to
$H$ reveals $H$ as the helper AS to all intermediate ASes on the $S \leadsto H$
half path, but $S$'s identity is hidden from all but the source AS, as the
route back to $S$ is stored as encrypted segments in the setup-packet headers.
$S$ uses the half-path to send to $H$ the final destination $D$ encrypted using
the public key of $H$, and then $H$ runs a back-off procedure to identify a
\emph{midway} AS $M$ who serves a similar role to Dovetail's dovetail AS.  The
back-off procedure will choose the midway to be the last AS on the $S \leadsto
M$ path who can transit traffic from his predecessor AS to $D$ without
violating any valley-free routing assumptions.  $M$ builds the second half-path
$M \leadsto D$ establishing the final end-to-end path $S \leadsto M \leadsto
D$.  So, ASes on a PHI path learn (1) the identities of their immediate
predecessor and successor ASes on the path, (2) their relative position on path
(\ie{} before the midway, the midway, or after the midway), and (3) the host/AS
at the end of their half-path (ASes on $S \leadsto H$ learn $H$'s identity,
ASes on on $M \leadsto D$ learn $D$'s identity).

{\bf Attack.} We run a modification of our
Section~\ref{subsec:dovetail_multiple_conn} \sysname{} attack on PHI.  In this
attack, the adversary compromises a single, fixed AS and attempts to
deanonymize a client who is repeatedly connecting to a fixed destination using
many helper nodes.  Path usage leaks location information in PHI just as it
does in Dovetail and so the adversary, when having compromised an AS on a
client's path, can use the topological information he learns to infer the
client's location.  When the adversary is on the $M \leadsto D$ half-path (and
therefore knows $D$'s identity), we suppose the adversary can link the
connection and his observation to the client using one of the linking
techniques described in Section~\ref{sec:models}.

We codify the adversary's observations, notated as $O$, as triplets
containing (1) the adversary's predecessor AS in the path, (2) the adversary's
relative position on path, and (3) the AS containing $D$.  Having made
observations $(O_1, \dots, O_n)$ the adversary computes
posterior probability $\Pr(L \mid O_1, \dots, O_n)$ for
each possible client location $L$, assuming a uniform prior.  Unlike Dovetail,
computing observation likelihoods in PHI is not prohibitive and so we favor a
probabilistic approach; however, we conservatively assume the adversary does
not know how often \emph{non-observations} occur, \ie{} when the client makes a
connection \emph{not} containing the compromised AS.  Because of our
conservative approach, the adversary may incorrectly compute some observation
likelihoods, leading to an incorrect posterior, and so we cannot accurately use
entropy to measure anonymity.  Instead, we use the adversary's posterior
probabilities as \emph{guessing scores} and use accuracy and rejection rates to
measure attack efficacy, similar to our Section~\ref{subsec:hornet_mobility}
HORNET mobility attack.

\textbf{Methodology.}
We run our \sysname{} attack in a simulated PHI network according to the
protocol description above.  This simulation is performed on a well-connected
AS graph containing \numPhiASes{} ASes.  The graph is generated using
shortest-path, valley-free inference using CAIDA's AS relationships.

We note that Chen \etal{} suggest that PHI clients select helper ASes for a
connection by attempting to maximize source-anonymity-set sizes with respect to
the connection's midway node; however, for two reasons, we do not implement
this helper selection scheme. First, in that proposal, it is also suggested
that clients only use subsets of helper ASes when maximizing anonymity-set
sizes to prevent an adversary from learning too much by observing a
client's helper selection, but it is unspecified how clients choose
appropriate helper subsets. Second, the method of anonymity-set computation
used in PHI does not properly model an adversary who can
reason about the actions of clients.  Suppose that, in some path,
midway AS $M$ observes predecessor AS $P$.  PHI describes a possibilistic
method for anonymity-set computation, in which $M$ would include any AS
with a valley-free path to $P$ in the source's anonymity set; however, if,
for example, all but a few source ASes use $P$ with negligible probability,
an adversary may be able to determine the source's AS confidently by way of
statistical inference.  Correctly maximizing anonymity-set sizes through
helper selections in the presence of this adversary (without leaking additional
topological information) is a non-trivial task and outside the scope of this
work.

\textit{Midway Frequency.}
To choose a compromised AS for this attack, we identify ASes likely to be
chosen as the midway in PHI connections.  The midway is the closest on-path AS
to the source who learns the destination of the connection, and so ASes who
frequently serve as the midway are well-positioned to run our linking attack.
We draw (source, helper, destination) AS triples uniformly at random from our
network and build a connection from the source to the destination via the
helper, recording which AS is selected to serve as the midway.  We repeat this
sampling \commanum{10000} times.

\textit{Anonymity Evaluation.}
We run simulations to measure how much a single compromised AS can learn in the
repeated-connections setting.  For each simulation, we (uniformly at random)
sample a source AS, a destination AS, and 500 ASes to serve as helper nodes.
We simulate multiple PHI connections from the source to destination --- for
each connection, the source chooses a helper uniformly at random (with
replacement) from among the 500 available to him.  If the connection passes
through the compromised AS such that the compromised AS learns the destination,
we compute the (possibly incorrect) likelihoods that this connection was
created by a client in each of the \numPhiASes{} possible source ASes.  For
each observable connection, we use these likelihoods to update the adversary's
belief about the location of the client (maintained persistently for the entire
simulation).  The adversary starts with a uniform prior belief over all
possible source ASes in each simulation.  For a single simulation, we consider
up to 50 repeated connections from the sampled source to destination.  We
repeat this entire simulation procedure \commanum{1000} times.

\begin{figure}[t]
   \centering
   \includegraphics[width=\columnwidth]{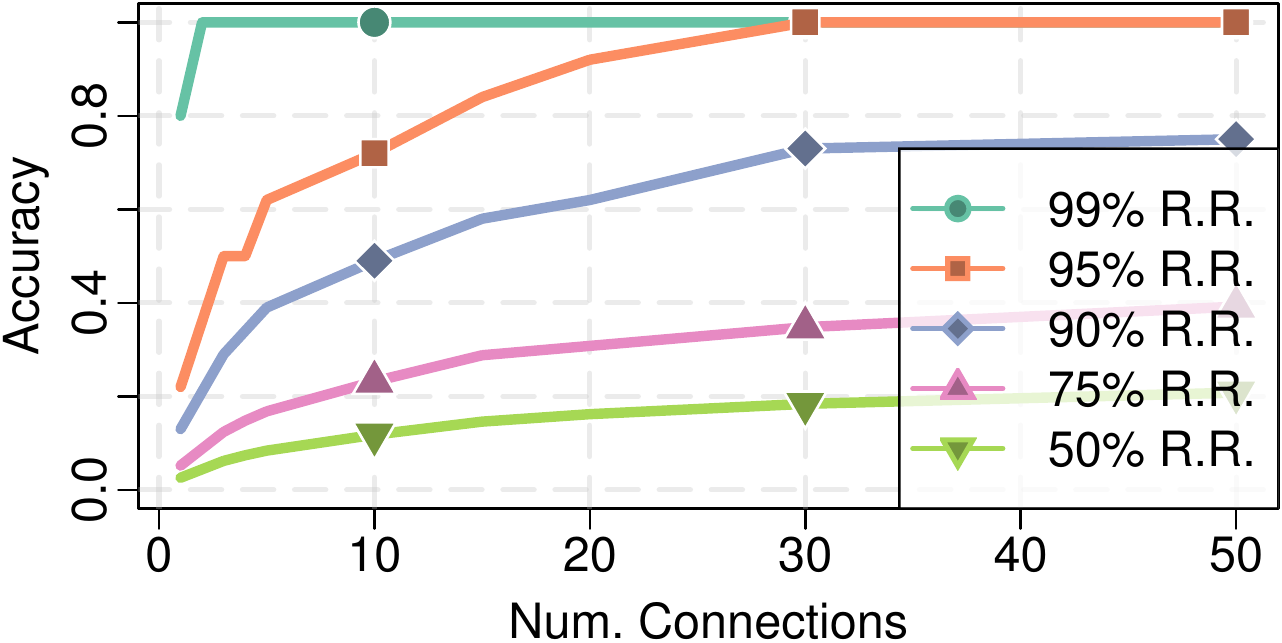}
   \vspace*{-.2in}
   \caption{Accuracy of deanonymization in PHI after a client makes $x$
   repeated connections for various rejection rates (R.R.) with respect to
   adversarial AS174.}
   \vspace*{-.2in}
   \label{fig:phi}
\end{figure}

\textbf{Results.}
\textit{Midway Frequency.}
AS174 (Cogent) is most frequently selected, serving as the midway in $7.9\%$ of
all connections; so, we use AS174 as our compromised AS for this attack.

\textit{Anonymity Evaluation.}
Figure \ref{fig:phi} plots the deanonymization accuracy the adversary can
achieve after the client makes some number of repeated connections at various
rejection rates.  For example, the point at $(x = 30,\ y = 0.35)$ in the 75\%
rejection rate line indicates that after the client made 30 repeated
connections, the adversary could determine the client's location in 25\% of
samples with 35\% accuracy (the adversary makes no guess due to lack of
confidence in the other 75\% of samples).

We find that PHI exhibits weaknesses to our \sysname{} attack.  As a client
continues to make connections through the network, the adversary's ability to
correctly determine the client's location grows significantly.  When the
adversary is willing to guess for 10\% of samples (\ie{} at a 90\% rejection
rate), he only achieves 13\% accuracy after a single connection; however, by 30
repeated connections, accuracy increases significantly to 73\%.  If the
adversary wishes to avoid incorrect guesses, he can tune his guessing threshold
to 99\% rejection and can achieve 100\% accuracy after just 2 repeated client
connections.

\end{document}